# Towards using Cough for Respiratory Disease Diagnosis by leveraging Artificial Intelligence: A Survey

Aneeqa Ijaz*, Muhammad Nabeel*, Usama Masood*, Tahir Mahmood*, Mydah Sajid Hashmi†,
Iryna Posokhova‡, Ali Rizwan§, and Ali Imran*
*AI4Networks Research Center, Dept. of Electrical & Computer Engineering, University of Oklahoma, USA
†University of Pittsburgh Medical Center, USA, ‡Kharkiv National Medical University, Ukraine
§Department of Electrical Engineering, Qatar University, Qatar
Email: aneeqa@ou.edu

*Abstract*—Cough acoustics contain multitudes of vital information about pathomorphological alterations in the respiratory system. Reliable and accurate detection of cough events by investigating the underlying cough latent features and disease diagnosis can play an indispensable role in revitalizing the healthcare practices. The recent application of Artificial Intelligence (AI) and advances of ubiquitous computing for respiratory disease prediction has created an auspicious trend and myriad of future possibilities in the medical domain. In particular, there is an expeditiously emerging trend of Machine learning (ML) and Deep Learning (DL)-based diagnostic algorithms exploiting cough signatures. The enormous body of literature on cough-based AI algorithms demonstrate that these models can play a significant role for detecting the onset of a specific respiratory disease. However, it is pertinent to collect the information from all relevant studies in an exhaustive manner for the medical experts and AI scientists to analyze the decisive role of AI/ML. This survey offers a comprehensive overview of the cough data-driven ML/DL detection and preliminary diagnosis frameworks, along with a detailed list of significant features. We investigate the mechanism that causes cough and the latent cough features of the respiratory modalities. We also analyze the customized cough monitoring application, and their AI-powered recognition algorithms. Challenges and prospective future research directions to develop practical, robust, and ubiquitous solutions are also discussed in detail.

## I. INTRODUCTION

Over the course of recent years, the healthcare domain has experienced a myriad of refinements in terms of cutting edge technologies and innovative treatment methods; notwithstanding, tremendous efforts are still required to prioritize the notion of healthcare over sick care. By following the state-of-the-art retrospective practices, when the patients provide their medical history in the clinical settings, they might not be discrete about reporting the symptoms specifics. This continual monitoring of symptoms can cause predicaments; worse case scenarios may lead to the diagnostic errors. Therefore, for the eradication of this issue, there have been increasing efforts towards the development of predictive and representative healthcare diagnosis systems. The revolution of existing healthcare practices is possible if we redirect the research focus towards preventive continuous monitoring instead of continual monitoring of symptoms, hence, prioritizing healthcare over sick care and leaving disease treatment as a last resort.

Due to the unprecedented interest towards data-driven processes and intelligent software, Computer Aided Detection (CAD) and Artificial Intelligence (AI)-based tools are gaining much attention. These intelligent data-driven medical models are showing considerable potential in assisting radiologists, healthcare professionals, and medical practitioners for the patient examination and accurate diagnosis, thus revolutionising a phenomenal and integrated healthcare sector [1], [2]. Thus, Artificial Intelligence (AI) is poised to play a significant role to make the healthcare system more cost effective, personalized, precise, and proactive [3]–[5].

Due to the prodigious improvements in the computing power and storage resources in the last decade, AI has already gained substantial attention in many fields including smart healthcare [6]–[8]. In response to the increased importance of using non-intrusive intelligent methods, and with the advent of AI, techniques have been evolved to become more refined and automated for the efficient information extraction from imaging modalities such as magnetic resonance imaging (MRI), computed tomography (CT), and ultrasound, in order to ensure better patient care. In addition, several Machine Learning (ML)-based frameworks are being leveraged for the general diagnosis of virulent maladies related to various systems, for instance, neurological [9], cardiovascular [10], digestive [11], and respiratory systems [12]. Although AI algorithms have already demonstrated acclaimed performance in image-based diagnosis [13], [14], voice-based diagnosis is also gaining attention and making prominent progress as sounds carry signature of numerous diseases [15]–[17]. For example, it has been shown that respiratory sounds, when analyzed by leveraging ML or Deep Learning (DL) techniques, can provide significant insights, thus enabling a powerful diagnostic tool [18]–[22]. These AI models demonstrate highly accurate and predictive outcomes, and can play an indispensable role in revitalizing the healthcare practices.

Among other respiratory sounds like wheeze, crackles, breathing, and stridor, cough is one of the outstanding sounds exhibiting unique features. Cough contains vital information for numerous respiratory conditions [23] manifesting as a symptom of over twenty medical conditions. The distinct latent features in cough can be exploited for the detection as well as preliminary diagnosis of various diseases by leveraging sophisticated and intelligent ML or DL algorithms trained through the cough acoustic data. Since the start of this century, there is a rapidly moving research landscape in this particular domain, several AI-based solutions have been presented which successfully detect cough events in the presence of other environmental noise and diagnose related respiratory diseases [24]. The feasibility of diagnosing numerous respiratory diseases with satisfactory high accuracy leveraging cough is supported by many studies [25]–[31]. In [31], researchers showed that cough alone has the potential to be used as a diagnostic tool to classify diseases such as, asthma, pneumonia, bronchiolitis, croup, and lower respiratory tract infections with over 80% sensitivity and specificity. Recently, viability of an AI-based approach analyzing cough sounds for the diagnosis of COVID-19 is demonstrated in [32]. Mostly, these AI-based algorithms run directly on portable devices such as smartphones and smartwatches, hence triage screening of pulmonary diseases can be performed free of charge at home. Nonetheless, the main focus of these works is the detection or diagnosis of different lung conditions (e.g., pulmonary nodules, tuberculosis, and interstitial lung diseases) in chest radiography with the aid of AI. A systematic review for the computer-based analysis of lung auscultations and respiratory sounds is also presented in [33]. In the literature, efforts have also been made in collecting and presenting works that deal with pulmonary diseases using AI [34], [35]. Recently, to overview the existing manual and automatic settings for cough counting and highlight the popular signal processing and AI techniques used in the cough monitoring devices, the authors presented a short review in [36]. Another article [37] provided a brief review of DL classification models for the pulmonary diseases leveraging cough audio analysis, the authors also provided the diagnosis of ten respiratory conditions. The outbreak of ongoing COVID-19 pandemic steered the focus of research to investigate the potential of AI and the advances of ubiquitous computing for timely diagnosis and to combat the spread of the disease [32], [38]–[45].

To provide a comprehensive and systematic overview of all the ongoing research efforts particularly in the domain of AI and ML by leveraging speech, image, and textual data for timely diagnosis of COVID-19, there are some notable studies [46]–[50]. In [46], the authors provided an exhaustive list of open access databases including CT scans, X-ray images, text data, and cough sounds for COVID-19 diagnosis. A systematic review is also conducted around six key questions discussing the usability of AI/ML methods for disease classification, prediction, risk assessment, and vaccine development for COVID-19 [48]. The authors briefly mentioned some speech and audio datasets. A comprehensive survey is conducted to study the five important use cases of AI for COVID-19 including COVID-19 diagnosis by leveraging data (images, sound, and text), prediction, patient behaviors, vaccine development and constructing supporting applications, and by exploiting machine learning techniques [47]. Other related works [49]–[51] review the ML and DL methods for COVID-19 diagnosis using medical images, non-invasive measures, and sound acoustic analysis for several applications such as timely treatment, emotion detection, and disease spread surveillance. However, the scope of these studies is towards the ongoing pandemic, i.e., COVID-19, by leveraging the available datasets.

There is a need for an exhaustive review that captures and discusses the potential of cough-based data-driven AI models for the diagnosis of numerous respiratory diseases, discussed in the literature. Therefore, in this work, we provide a detailed survey of techniques that have been presented in the literature for cough detection and diagnosis of different diseases using cough within one framework. In particular, we present a comprehensive survey of the existing literature on cough-based ML or DL models for the detection and diagnosis of over 25 respiratory conditions. The main focus is to summarize all the important algorithms discussed in the literature to date that offer a reliable accuracy in predicting a disease by just using the cough sounds. Thus, encouraging the researchers to find the information easily while working in the healthcare sector.

Our main contributions of this work are summarized as follows:

- We delineate the mechanism that produces cough, the distinctive features, and the cough types that are considered as hallmarks of certain diseases. We provide the motivation of using CAD and ML tools for the characterization of subtle/nuance features in cough for better diagnosis.
- We identify the common respiratory conditions and characterize the types of cough based on duration and compare the cough characteristics.
- We provide the important procedures that are essential for the development of a robust ML/DL-based framework for efficient and accurate detection/diagnosis. This includes the data acquisition, preprocessing, feature engineering, and model training.
- We also summarize the state-of-the-art AI-based solutions that exploit unique latent cough features to successfully detect, monitor, and diagnose different respiratory diseases.
- We provide a brief overview of the detection and diagnosis ML techniques trained on the non-cough sounds.

*Organization of the Paper:* The important acronyms used in this article are summarized in Table I, whereas Fig. 1 shows the taxonomy of the sections and subsections discussed in this study. The next Section II discusses the mechanism that produces cough and its distinctive features. Section III presents the types of cough based on duration, while Section IV describes the data acquisition methods. ML/DL pipeline in detection/diagnosis for cough based respiratory conditions is explained in Section V. Section VI briefly discusses about the detection and diagnosis leveraging non-cough sounds. Different open research issues and future recommendations are outlined in Section VII. Finally, we conclude the paper in Section VIII.

TABLE I. LIST OF ACRONYMS

| Acronym | Description |
|---|---|
| AI | Artificial Intelligence |
| ADAM | Automated System for Asthma Monitoring |
| ANN | Artificial Neural Network |
| BC | Bayes Classifier |
| CNN | Convolutional Neural Network |
| CPNN | Constructive Probabilistic Neural Network |
| DBN | Deep Belief Network |
| DT | Decision Tree |
| DL | Deep Learning |
| DWT | Discrete Wavelet Transform |
| DNN | Deep Neural Network |
| ECC | Energy Cepstral Coefficients |
| FT | Fourier Transformation |
| FCM | Fuzzy C-means Clustering |
| GFCC | Gammatone Frequency Cepstral Coefficient |
| GMM | Gaussian Mixture Model |
| GMM-UBM | Gaussian Mixture Model–Universal Background Model |
| HACC | Hull Automatic Cough Counter |
| HMM | Hidden Markov Models |
| k-NN | k Nearest Neighbor |
| LCM | Leicester Cough Monitor |
| LSTM | Long Short Term Memory |
| LPCS | Linear Predictive Coding Spectrum coefficients |
| LVQ | Learning Vector Quantization |
| LR | Logistic Regression Model |
| LPCC | Linear Prediction Cepstral Coefficients |
| LPC | Linear Prediction Coefficients |
| LSF | Line Spectral Frequencies |
| LogE | Log Energy |
| MelSpec | Melscaled spectrogram |
| MFCC | Mel Frequency Cepstral Coefficients |
| MR | Multiple Regression |
| MLP NN | Multilayer perceptron Neural Network |
| MFB | Mel Filter Bank Dimension |
| ML | Machine Learning |
| NN | Neural Network |
| NB | Naive Bayes |
| NNC | Nearest Neighbor Classification |
| NMF | Non-negative Matrix Factorisation |
| ONN | Octonionic Neural Network |
| OC-SVM | One Class SVM |
| PNN | Probabilistic Neural Network |
| PLP | Perceptual Linear Prediction |
| PCA | Principal Component Analysis |
| RF | Random Forest |
| RNN | Recurrent Neural Network |
| SVM | Support Vector Machine |
| UBM | Universal Background Model |
| Zcr | Zero Crossing |

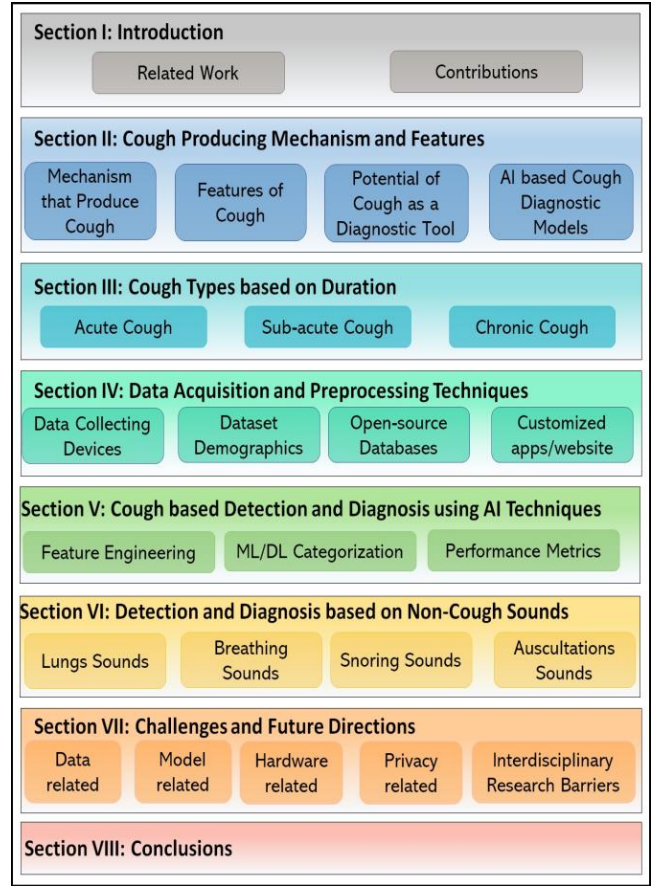

Fig. 1: Organization of the paper.

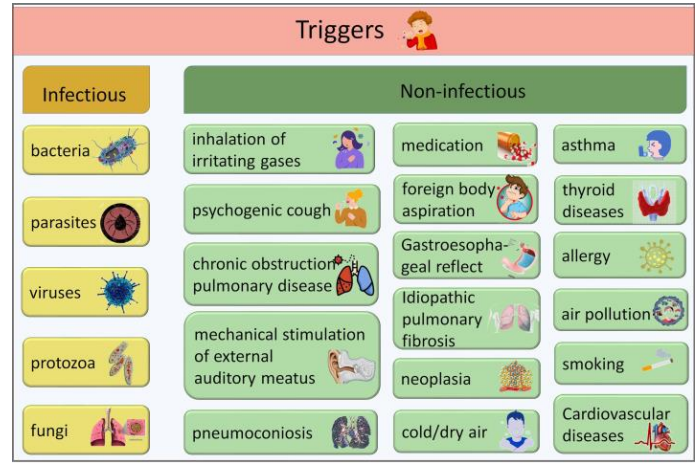

Fig. 2: A taxonomy of cough triggers.

## II. Cough Producing Mechanism and Features

Cough is one of the most common signatures of numerous respiratory diseases. It is usually caused by either viral or bacterial respiratory infections. Fig. 2 shows the taxonomy of infectious and non-infectious triggers that manifest cough as a symptom and have the possibility of eventually leading to a medical condition. In this section, we show how different respiratory conditions associated with different diseases have distinctive cough acoustics. These distinguishable cough features provide an opportunity for the AI models to get training and identify the underlying disease in a timely manner, thus providing assistance to the medical professionals.

### A. Mechanism that Produces Cough

Cough is a protective reflex mechanism against any foreign object or irritant. However, it becomes a pathological reflex when it does not perform the function of clearing the airways. In some cases, this mechanism causes pain in the throat, chest, or behind the sternum. In addition, it has the ability to affect not only the psychosocial domain but is potentially harmful to the patient's airway mucosa [52].

Cough reflex consists of afferent, central, and efferent components. It is initiated by the stimuli of vagus nerve which functions as cough receptors [53]. Cough receptors are spread through pharynx, larynx, trachea, main carina, bronchi, and smaller distal airways. In the literature, different types of coughs can be found. Fig. 3 provides a thorough illustration of the types of cough related to different diseases, and their association with the various anatomical sites of the respiratory system. The receptors located on the larynx, pharyngeal wall, and tracheobronchial tree respond to mechanical stimuli and various chemical irritants. The external auditory canal, eardrum, paranasal sinus, diaphragm, pericardium, pleura, and pericardium have the mechanical receptors. The airway afferent nerves include: rapidly adapting receptors, slowly adapting stretch receptors, unmyelinated C fibers, and $A\delta$ cough fibers. Rapidly adapting receptors include $A\beta$ stretch receptors which have a low threshold for activation by stimuli. The rapidly and slowly adapting stretch receptors preferentially respond to mechanical relative to chemical stimuli. They respond to various mechanical stimuli including variation in lung volumes, airway smooth muscle constriction, and pulmonary edema. These receptors are relatively insensitive to the majority of direct chemical stimuli but respond to the mechanical changes induced by chemical stimuli including changes in lung volumes, airway smooth muscle tone and mucous hypersecretion. For example, chemical stimuli like capsaicin and bradykinin induce smooth muscle contraction, mucous secretion, and fluid accumulation in lung parenchyma that activates RARs and SARs [54]. C fibers have lower sensitivity than rapidly and slowly adapting receptors for mechanical stimuli. There are likely two or more types of the involuntary cough reflex. The *first type* is caused by the stimulation of bronchopulmonary C-fibers by inflammation, tissue injury, or chemical irritants (including capsaicin, sulfur dioxide, bradykinin, and citric acid) and associated with sensation "urge to cough". The *second type* aspiration-induced involuntary cough reflex occurs with the activation of mechanosensory-responsive $A\delta$ fibers [55]. On stimulation, the cough receptor sends afferent impulses via the vagus nerve to the cough center in the medulla and interconnect to neural networks in the brain. The cough center creates a signal that propagates to the vagus, phrenic, and spinal motor nerves to expiratory musculature to induce coughing [56].

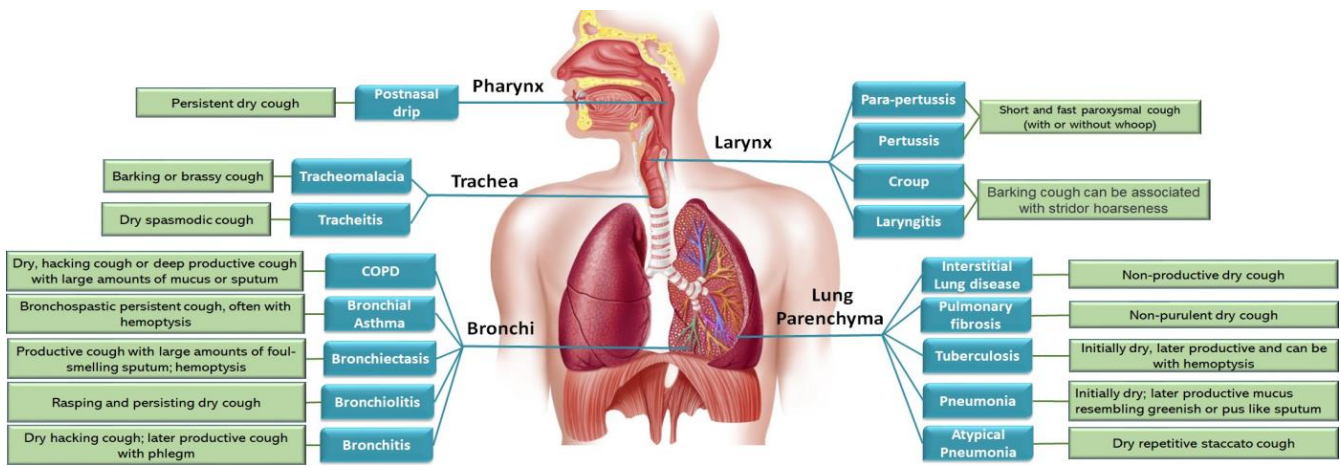

Fig. 3: Types of cough.

### B. Features of Cough

There are several types of cough with unique characteristic features of sound which healthcare professionals are able to distinguish. However, there are many respiratory conditions that can cause same type of cough, making it difficult to be distinguishable by the human ear. The eminent five cough features that help the physicians to identify the associated diseases are:

- *Dry cough*: It is also referred to as a non-productive cough because it does not produce phlegm and sounds like a hacking cough. It can be caused, for example, by irritants such as air pollutants, non-infectious conditions such as upper airway cough syndrome, asthma, heart failure, gastro-esophageal reflux disease, side effects of some medications, or viral infections of upper respiratory tract [57].
- *Wet (productive) cough*: This type of cough produces phlegm, sometimes with impurities of pus or blood; sounds low, heavy, mucousy, and may come with a rattling or wheezing sound as well as tightness in the chest [58]. Productive cough is often caused by infectious diseases such as flu, cold, bronchitis, pneumonia, tuberculosis, lung abscess, or other conditions including bronchiectasis, chronic obstructive pulmonary disease, and cystic fibrosis.
- *Whooping cough*: Spasmodic and continuous coughing, which result in intense inhalation after the episode and produce the whooping sound. This cough can be an indication of pertussis [59].
- *Barking cough*: It sounds like a barking seal, often with a stridor (high-pitched whistling sound during inhalation or exhalation). Barking cough is a symptom of croup, tracheomalacia or psychogenic cough [60].
- *Staccato cough*: Cough that comes with a series of outbursts having at least one breath in between two consecutive outbursts. It is caused by chlamydia pneumonia in infants and is regarded as staccato due to inspiration between each single cough [61].

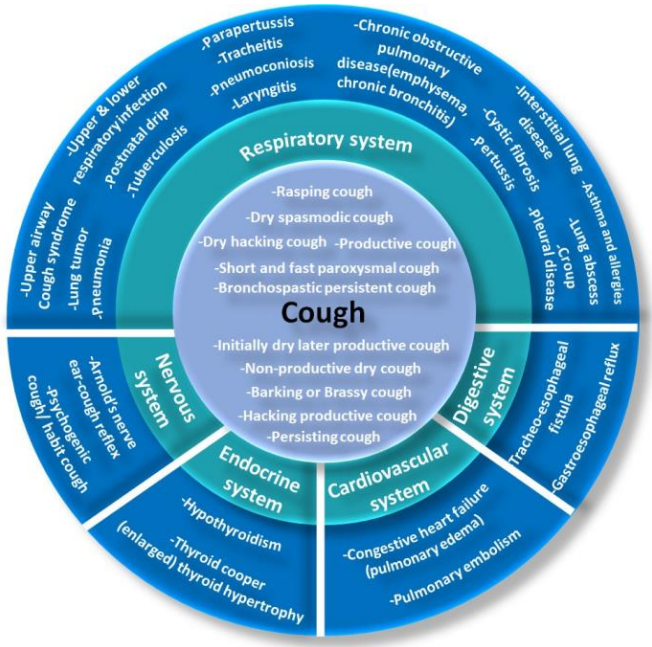

Fig. 4: Cough types and the associated diseases.

*C. Potential of Cough as a Diagnostic tool*

Cough is considered as a symptom of various ailments. Experienced healthcare professionals use cough reflex interpretation as one of the manifestations helping them to associate it with discrete diseases. For instance, barking cough is associated with pertusius, whereas common causes of dry or hacking cough are upper airway cough syndrome (UACS), asthma, and gastroesophageal reflux disease (GERD). In case of viral croup, inflammation of the subglottic soft tissue occurs resulting in a rasping quality of the voice. Fig. 4 shows a summary of cough types as well as different diseases associated to various human systems.

Trained and experience physicians have been exploiting cough signatures to perform a differential diagnosis for various respiratory conditions, for instance, COPD, pneumonia, bronchitis, pertussis, pharyngitis, asthma, and tracheitis [23], [62], [63]. The reason for the accurate diagnosis by using cough is possible because the symptoms and the location of the underlying irritants in the respiratory system are quite distinctive, which lead to the audibly distinguishable cough sounds. Nonetheless, an untrained human ear is incapable of characterising the coughs caused by the different diseases delineated in the figure. It is evident from the figure that there is an overlap of cough types and the associated diseases, making it difficult for the physicians to use cough as a preliminary diagnostic tool with guaranteed accuracy.

For the characterization of cough based on acoustic analysis and auscultation, a human hear is able to distinguish cough based on limited audible features such as timbre, loudness, pitch, and duration. These different features along with their attributes and description are listed in Table II. Other features that can aid the medical professionals to further classify the cough are wetness and dryness. Hence, common cough types that are considered as hallmark of certain diseases are identified based on the limited audible features. The fact that cough can be characterized in only five or six dimensional space by its virtue is limiting its ability to be uniquely associated with a certain disease. A detailed study is performed in [64] to analyze how the healthcare professionals interpret cough sounds based on the acoustic features. The results suggested that the medical professionals are able to recognise some of the qualities of the cough sounds based on the aforementioned limited features, however, the rate of accurate clinical diagnosis was identified only in 34% of cases. It is demonstrated in the study that dry or wet cough were precisely recognized in 76.1% of cases, whereas cough with accompanied wheeze was recognized only in 39.3% of cases, as it is difficult to distinguish wheeze in the cough [64]. Therefore, it can be interpreted that there is an overlap between the characteristics of different cough sounds, making it hard to be distinguishable by human ear based on the restricted features. Thus, it is ineffective to use cough for preliminary diagnosis as the limited audible features are not enough to delineate all the diseases. Based on these facts, it is difficult to project cough in finer characterization space for the detailed analysis. Consequently, this is a dire need to complement the existing medical practices and to provide additional assistance to the physicians for better cough characterization and accurate disease diagnosis.

TABLE II. FEATURES THAT HELP HUMAN EAR TO DIFFERENTIATE THE VOICE, GENERALLY USED FOR CLINICAL DIAGNOSIS

| Feature Name | Attribute | Description |
|---|---|---|
| Timbre | Harmonics, temporal | Multiple frequencies changing through time |
| Pitch | Psycho-acoustical (frequency) | The perception of a high or low sound |
| Quality | Temporal | Sensation received by the ear |
| Loudness | Amplitude | The intensity of a sound |
| Duration | Period | Length of time a pitch, or tone, is sounded |

*D. AI-based Cough Diagnosis models*

Building on the insights from the healthcare domain knowledge, it can be argued that it is possible to develop a robust, reliable, and efficient AI-powered screening and diagnosis tool using cough sounds. This is because ML and DL models have the ability to process huge datasets and analyze hundreds of hidden features of coughs, which are beyond human capability to comprehend [65]–[68]. By using these models, we can project cough in higher dimensional space and, hence, able to separate new unique features that are otherwise not possible to be distinguished by human ear due to the overlapping in the lower dimensional space [69]–[73]. For instance, a human ear can detect only a handful of features (i.e., four to five). These features have binary possibilities such as dry or wet cough, barking or no barking cough, etc. Thus, there is a possibility of only $2^5 = 32$ combinations of features.

In case of Artificial Intelligence, ML/DL models are able to classify numerous features, each feature further having a range of possible values. Theoretically, these models can classify between millions of possible combinations of features which human ear cannot distinguish. As shown in Table III, for each audio subframe, there exists at least 312 unique temporal, spectral, and statistical cough features that can be possibly leveraged to build a robust and efficient ML model for a cough-based diagnosis engine. It is also well known that AI has the potential to differentiate subtle distinctions and nuances in the cough that are associated with the unique diseases. Hence, it can be exploited as a test medium for the diagnosis of diverse respiratory diseases. Several independent studies [68], [74], [75] have backed up this claim and demonstrated that the distinct cough latent features can be used for the accurate AI-based diagnosis of the respiratory diseases.

To perform cough detection and preliminary disease diagnosis based on cough, we present a robust yet easy to develop AI engine in Fig. 5. The major phases of a cough detection/diagnosis framework are data acquisition, data pre-processing, feature extraction, and selection of the appropriate machine learning models. These phases and their relevant characteristics are elaborated in the subsequent sections. The main task of the AI engine is to run a cough detection test and analyze whether the recorded sound is a cough or not. Further, it should be robust enough to detect cough in the presence of background noise. In case of diagnosis, the aim for this AI engine is to analyze cough and diagnose the correct disease accurately. By delegating most of the processing to the cloud; such an AI engine can be installed on portable devices. Thus, making it a resourceful screening tool easily accessible to patients for monitoring themselves and in some cases, providing preliminary diagnosis in order to encourage the patients to seek timely medical assistance. Such intelligent diagnostic tools can also aid physicians in better planning and providing care, ultimately leading to better outcomes and increased patient satisfaction.

## III. COUGH TYPES BASED ON DURATION

Based on the duration, cough can be broadly classified into three categories: *acute* (less than three weeks), *subacute* (three to eight weeks), and *chronic* (longer than 8 weeks) [76], as shown in Fig. 6. Other details about symptoms and diagnostic methods are mentioned in Table IV. In this section, we explain in detail the diseases associated to the aforementioned types of coughs, relevant symptoms, and the existing medical practices for cough analysis. We also discuss how ML/ DL classifiers can complement the existing state-of-the-art healthcare procedures for timely and accurate diagnosis.

### A. Acute Cough:

Upper or lower respiratory tract infections and acute exacerbation of pre-existing conditions like chronic obstructive pulmonary disease (COPD), asthma, and bronchiectasis can cause acute cough. The acute attack of asthma and COPD can be clinically severe and sometimes life-threatening, .

i. *Upper Respiratory Tract Infections (URTI):* URTIs include common cold and croup. Common cold can cause upper respiratory tract inflammation through various viruses and has a self-limiting course of seven to ten days with dry or productive cough.

- *Croup:* Croup is usually a viral infection that also causes inflammation of the upper respiratory tract. It presents with abrupt onset of barking cough along with the inspiratory stridor and breathing difficulty [77], [78]. It is a clinical diagnosis (distinct barking cough), and viral antigen detection or serology is not recommended to confirm the diagnosis, as it is very expensive. Therefore, devising some innovative and cost-effective diagnosis methods by leveraging the tools of artificial intelligence can be of great help [79].

ii. *Lower Respiratory Tract Infections (LRTI):* These infections can be caused by a viral or bacterial infection. These infections include bronchitis, pneumonia, atypical pneumonia, and lung abscess. The infections can involve bronchial tree and lung parenchyma or both.

- *Acute Bronchitis:* In acute bronchitis, the main presenting complaint of the patients is cough, which is productive with green or clear phlegm along with dyspnea and wheeze [80]. The involved cough is usually self-resolving within three weeks. Acute bronchitis is a diagnosis of exclusion that is made after ruling out other respiratory diseases like pneumonia, acute exacerbation of COPD, and asthma. However, manual cough assessment can sometimes lead to misdetection because of the overlapping audible features.
- *Pneumonia:* It is the infection of lung parenchyma caused by bacteria, viruses, or fungi. Patients present with acute onset of productive cough, fever, tachycardia, and increased respiratory rate. Patients with symptoms suggestive of pneumonia have blood samples and culture drawn out before initiation of empirical antibiotics. Radiological studies can show patchy or a lobar consolidation in the lungs with pleural involvement. Pleural fluid, sputum, and blood samples are drawn out for biochemical tests and culture. Several studies have been carried out to compare the accuracy and efficiency of sputum polymerase chain reaction (PCR) and culture for the diagnosis of pneumonia [81]–[83]. They showed that sputum PCR is also more sensitive than culture in diagnosing community-acquired pneumonia in the hospitalized patients. In patients with bacteremia, PCR of blood sample is more sensitive than the culture in diagnosing pneumonia [84].
- *Atypical Pneumonia:* Atypical pneumonia presents with extrapulmonary findings in addition to the lung findings. The patient presents with low-grade fever, cough mostly 7TABLE III. LIST OF AUDIO FEATURES THAT CAN BE POSSIBLY LEVERAGED FOR BUILDING A SCALABLE AND ROBUST ML CLASSIFIER FOR COUGH-BASED DIAGNOSIS ENGINE

| Sr. No. | Feature Name | No. of Features | Type | Description |
|---|---|---|---|---|
| 1 | Zero Crossing Rate | 1 | Temporal | The rate at which signal changes its sign during the duration of a particular frame, e.g., positive to negative or vice versa. |
| 2 | Energy | 1 | Temporal | The sum of squares of the amplitude of the signal, normalized by its frame length. |
| 3 | Entropy of Energy | 1 | Temporal | Abrupt changes in a sub-frame can be measured from its entropy of energy. |
| 4 | Log Energy | 1 | Temporal | The log energy for every subsegment |
| 5 | Skew | 1 | Statistical/Temporal | Measure of the asymmetry of the probability distribution of the data segment |
| 6 | Kurtosis | 1 | Statistical/Temporal | Measure of the "tailedness" of the probability distribution of the data segment |
| 7 | Spectral Centroid | 1 | Statistical/Spectral | Indicates the center of mass/gravity of the spectrum. Approximately related to timbral "brightness |
| 8 | Spectral Mean | 1 | Statistical/Spectral | 1st spectral moment |
| 9 | Spectral Variance | 1 | Statistical/Spectral | 2nd spectral moment |
| 10 | Spectral Skewness | 1 | Statistical/Spectral | 3rd spectral moment |
| 11 | Spectral Kurtosis | 1 | Statistical/Spectral | 4th spectral moment |
| 12 | Spectral Spread (SSp) | 1 | Statistical/Spectral | Variance of the spectrum around the spectral centroid |
| 13 | Spectral Slope (SSl) | 1 | Statistical/Spectral | Rate of decrease of the spectral amplitude towards the high frequencies, calculated using Linear Regression |
| 14 | Spectral Crest Factor | 1 | Spectral | Ratio of the maximum spectrum power and its mean |
| 15 | Spectral Bandwidth | 1 | Spectral | Measure of the spectral dispersion |
| 16 | Spectral Flatness | 1 | Spectral | Measure of noisiness in a signal computed by the ratio of the geometric mean and arithmetic mean of its energy spectrum |
| 17 | Spectral Entropy | 1 | Spectral | Entropy of the normalized power spectral density of a signal |
| 18 | Spectral Flux | 1 | Spectral | Measure of the rate of change of power spectrum between two successive frames, calculated by the Euclidean distance between their normalized spectra |
| 19 | Bispectrum Score (BGS) | 1 | Spectral | Measure of nonlinear interactions in a signal, calculated by its third-order spectrum |
| 20 | Pitch | 1 | Spectral | The fundamental frequency of the audio signal |
| 21 | MaxF | 1 | Spectral | Maximum Frequency |
| 22 | Band Power | 1 | Spectral | Average power in the input signal |
| 23 | Spectral Rolloff | 1 | Spectral | The roll-off frequency below which 85% of the spectrum's energy is concentrated. |
| 24 | Spectral Turbulence | 1 | Spectral | Measure of variations in the spectral content of a signal |
| 25 | Mel-Spectrogram | 20 | Spectral | Mel-frequency spectrogram coefficients for 20 Mel-bands |
| 26 | MFCCs | 13 | Spectral | Mel Frequency Cepstral Coefficients gives a concise representation of the shape of spectral envelope |
| 27 | Delta MFCCs | 13 | Spectral | Delta-MFCC are the first-order derivative of MFCC used to measure speech rate |
| 28 | Delta Delta MFCCs | 13 | Spectral | Delta-Delta-MFCC are the second-order derivative of MFCC used to measure speech acceleration |
| 29 | Chromagram | 12 | Spectral | Measure of spectral energy w.r.t. the 12 pitch classes of western-type music |
| 30 | Chroma Deviation | 1 | Spectral | The standard deviation of the chroma vector containing 12 chroma coefficients. |
| 31 | Constant-Q chromagram | 12 | Spectral | Constant Q transform based Chroma Values |
| 32 | Chroma Energy Normalized Statistics | 12 | Spectral | Chroma Energy Normalized variant of Chroma values |
| 33 | Cochleagram | 20 | Spectral | Gamma-tone filter based variant of spectrogram |
| 34 | Linear Predictive Coefficients (LPC) | 20 | Spectral | A compressed representation of the spectral envelope of a signal |
| 35 | Linear Predictive Cepstral Coefficients (LPCC) | 20 | Spectral | Cepstral representation of LPC |
| 36 | Line Spectrum Pairs (LSP) | 20 | Spectral | Direct mathematical representation of LPC coefficients for added filter stability and efficiency |
| 37 | Discrete wavelet transform (DWT) | 20 | Spectral | Decomposes a signal into a set of wavelets |
| 38 | Continuous Wavelet Transform (CWT) | 20 | Spectral | Time-frequency representation of a signal by decomposing it into wavelets |
| 39 | Perceptual linear prediction (PLP) | 20 | Spectral | Perceptual linear prediction coefficients gives more weight to the perceptually important spectrum regions |
| 40 | Formant Frequencies (FF) | 4 | Spectral | Formant frequencies have high energy in the spectrum of a human speech signal |
| 41 | Non-Gaussianity Score (NGS) | 1 | Spectral | NGS gives the measure of deviation from Gaussianity of a signal |
| 42 | Power Spectral Density (PSD) | 20 | Spectral | Measure of signal's power distribution versus frequency. |
| 43 | Tonnetz | 6 | Spectral | Tonal centroid features (tonnetz) |
| 44 | Spectral Contrast | 7 | Spectral | Measure of difference between the peaks and valleys in the spectrum of a signal |
| 45 | Local Hu Moments | 13 | Spectral | Measure the degree of how the energy is concentrated to the center of energy gravity of local region of spectrogram |

TABLE IV. CLINICAL PICTURE AND DIAGNOSTIC APPROACH FOR DIFFERENT TYPES OF COUGH

| Type of Cough | Reference | Disease | Clinical Picture | Diagnosis |
|---|---|---|---|---|
| **Acute cough** | [85], [86] | URTI | Nasal discharge, stuffiness, post nasal drip, sore throat | Mainly a clinical diagnosis. Nasopharangeal swab is done for antigen detection by ELISA, IFA |
| | [78] | Croup | Occurs mostly in children. Presents with catarrhal symptoms, barking cough, stridor and breathing difficulty | Diagnosed clinically |
| | [87], [88] | Acute bronchitis | Cough (dry/productive) can be preceded by nasal congestion, flu, sore throat, and headache. Rhonchi and wheeze may be heard on chest auscultation | Clinical diagnosis, Chest X-ray is usually normal or may show subtle changes like thickening bronchial walls in lower lung zones |
| | [89], [90] | Atypical Pneumonia | Prodromal symptoms of malaise, low-grade fever, myalgia, sore throat, flu-like symptoms. It is followed by a dry or productive cough, accompanied by pleuritic chest pain and dyspnea | Clinical diagnosis, molecular tests like nucleic acid amplification and PCR are reserved for clinically severe diseases with complications like hemolysis and mucocutaneous manifestations |
| | [91] | COVID-19 | Present with wide clinical symptoms including cough, myalgia, fever, headache, sore throat, shortness of breath, loose stools, chest discomfort, loss of taste, and smell sensation | The nucleic acid amplification test (NAAT) by reverse transcriptase-polymerase chain reaction (RT-PCR) has high sensitivity and specificity for detecting SARS-CoV-2 |
| | [84] | Pneumonia | Productive cough, 'rusty' colored sputum in streptococcus pneumonia, fever, tachycardia, tachypnea, and chest pain. Bronchial breath sounds, tactile and vocal fremitus on chest auscultation | Chest Xray, HRCT, sputum microscopy, PCR and blood culture |
| | [92]–[94] | Lung abscess | Clinical picture similar to pneumonia, cough initially dry but productive when abscess develops communication with bronchus | Chest X ray, HRCT shows walled-off abscess with air-fluid level, Brochoscopy for microbiolgical testing and biopsy |
| **Subacute cough** | [95], [96] | Pertussis | Catarrhal stage followed by bouts of productive cough with post-tussive emesis. | PCR and culture of sputum samples on Bordet-Gengou and Regan-Lowe agar |
| **Chronic cough** | [97] | Asthma | Nocturnal cough, breathlessness, wheezing and chest tightness. Symptoms triggered by exposure to allergen | Spirometry: FEV1/ FVC ratio< 70%, reversible by inhalation of beta agonists. Reversal and improvement of symptoms by inhalation of Beta 2 agonists. |
| | [65] | GERD | Dyspepsia, heartburn, metallic taste, cough, hoarseness of voice | 24 hours esophageal ph monitoring |
| | [77] | UACS | Dry cough, itching, post-nasal drip, sore throat, nasal stuffiness, and rhinitis | Clinical diagnosis |
| | [67] | COPD | Productive cough, dyspnea, common in smokers, pink puffers, blue bloaters | Spirometry: FEV1/ FVC ratio< 70%, not reversible by bronchodilators. Chest X ray. |
| | [98]–[100] | Tuberculosis | Productive cough, hemoptysis, night sweats, fatigue, weight lost, low-grade fever | Microscopy and culture from sputum smear, BAL washing, pleural fluid. NAA test on sputum. Chest Xray and HRCT are imaging studies of choice |
| | [101] | Congestive cardiac failure | Dyspnea on exertion and in later stages at rest, orthopnea, swelling in legs, cough with straw-colored sputum, basal crepitus on ascultatation at early stages | Echocardiography, chest X ray, BNP levels, cardiac stress test |
| | [102] | Idiopathic pulmonary fibrosis | Bouts of non-productive cough, dyspnea. Presents with systemic symptoms like uveitis, blurred vision, artharalgia, dyspepsia, dysphagia, rash. | Chest X-ray, HRCT, lung biosy and respiratory studies are carried out. |
| | [103] | Psychogenic cough | Repetitive cough in the absence of clinical disease. | Diagnosed after clinic and psychiatric evaluation. |
| | [104] | Cystic fibrosis | Genetic disorder with early childhood onset. Productive cough with wheeze, chest tightness, steatorrhea, poor weight gain. | New-born screening followed by sweat chrolide test. Nasal potencial difference test is not done routinely. |
| | [105] | Lung cancer | Cough, hemoptysis, weight lost, and fatigue | Sputum cytology, chest X-ray, HRCT, bronchoscopy and lung biopsy. Lumph node biopsy, screening for metastatic sites. |
| | [65] | Cough due to medications | Dry or productive cough | Improvement in symptoms in one to four weeks after discontinuation of medication suspected of resulting in cough. |

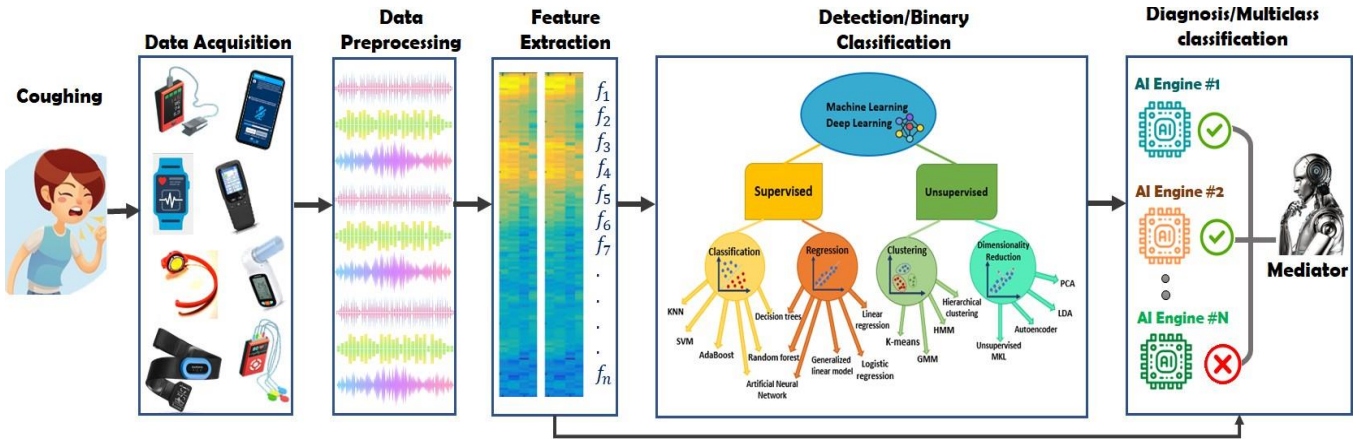

Fig. 5: AI-based cough detection/diagnosis process.

dry, dyspnea, wheeze, and myalgia. It is usually preceded by upper respiratory symptoms like rhinitis, sore throat, headache, and sinus pain [106]. In children, the dry repetitive staccato cough is characteristic of chlamydial pneumonia.

- *Lung abscess:* Abscess is walled off, well-circumscribed collection of necrotic tissue and pus. When the abscess develops communication with the bronchus, the cough becomes productive [92], [93]. Radiological tests like chest X-ray and high resolution computed tomography (HRCT) are needed to carry out to grasp the extent and dimensions of an abscess. However, there is a possibility that radiologists may misinterpret diseases because of inexperience or human error, leading to a misdiagnosis, i.e., false-negative result or false-positive result.
- *COVID-19:* Novel virus SARS-CoV-2 presents a wide clinical spectrum. It has variable clinical severity ranging from the majority of cases presenting with mild self-resolving course to critical fatal illness [107]. The common symptoms include cough, myalgia, fever, headache, sore throat, shortness of breath, loose stools, chest discomfort, and loss of taste and smell sensation [91]. Most of the time, viral illness affects the upper respiratory tract, but sometimes it can affect the lungs resulting in pneumonia with bilateral ground-glass infiltrates [108]. The nucleic acid amplification test (NAAT) by reverse transcriptase-polymerase chain reaction (RT-PCR) has high sensitivity and specificity for detecting SARS-CoV-2. The method employs the detection of two or more genomes, including envelope, spike, nucleocapsid, and RNA-dependent RNA polymerase [109].
- *Aspiration of the foreign body:* Another cause of acute cough is the aspiration of a foreign body, which can cause sudden onset of cough and it mostly occurs in children. Usually, the patient presents with a history of choking, followed by classical triad cough, wheeze, and decreased breath sounds. A chest radiography is an initial test performed [66]. Flexible bronchoscopy is both diagnostic and therapeutic. Automated AI-based techniques can make the diagnosis of such cough very secure, straightforward, and fast.

However, there are chances of misdiagnosis with the existing practices in the image analysis. According to statistics, the misdiagnosis rate caused by a human can reach up to 10–30% [110]. Therefore, there is an increasing trend of amalgamation of CAD systems to provide aid and helpful tools for the health professionals for accurate and efficacious diagnosis [35]. Machine learning and deep learning have been a vibrant area in AI for the promising results in the healthcare domain. These methods are considered as a powerful tool in the automatic detection of the disease from the datasets constitute of CT scan [111], X-Ray images [112], and respiratory sound data [113]. Thus, AI has the potential to provide assistance to the physicians by deploying rapid and low cost, yet accurate screening tools.

### B. Sub-acute Cough:

Sub-acute cough lasts for three to eight weeks. It is commonly due to post-infection and exacerbation of asthma, COPD, and upper airway cough syndrome [23]. Post-infection cough mostly occurs after viral infection and usually resolves within eight weeks. It occurs due to the increased sensitivity of the larynx. Mycoplasma pneumonia and pertussis can also result in post-infection cough [114].

- *Pertussis:* It is characterized into three clinical stages: (i) catarrhal, (ii) paroxysmal, and (iii) convalescence. The catarrhal stage has symptoms similar to the common cold. The paroxysmal stage is detected by bouts of whooping cough with post-tussive emesis and the cough is productive [95]. During the convalescent stage, the cough gradually recovers, and it can last from several weeks to months. There can be acute episodes of cough during this phase due to the superimposed URTIs [115]. The gold standard test for diagnosis of pertussis is culture. Nowadays, PCR has replaced culture as a test of choice [13]. However, an automated ML-based cough detection method can expedite the accurate and reliable diagnosis.

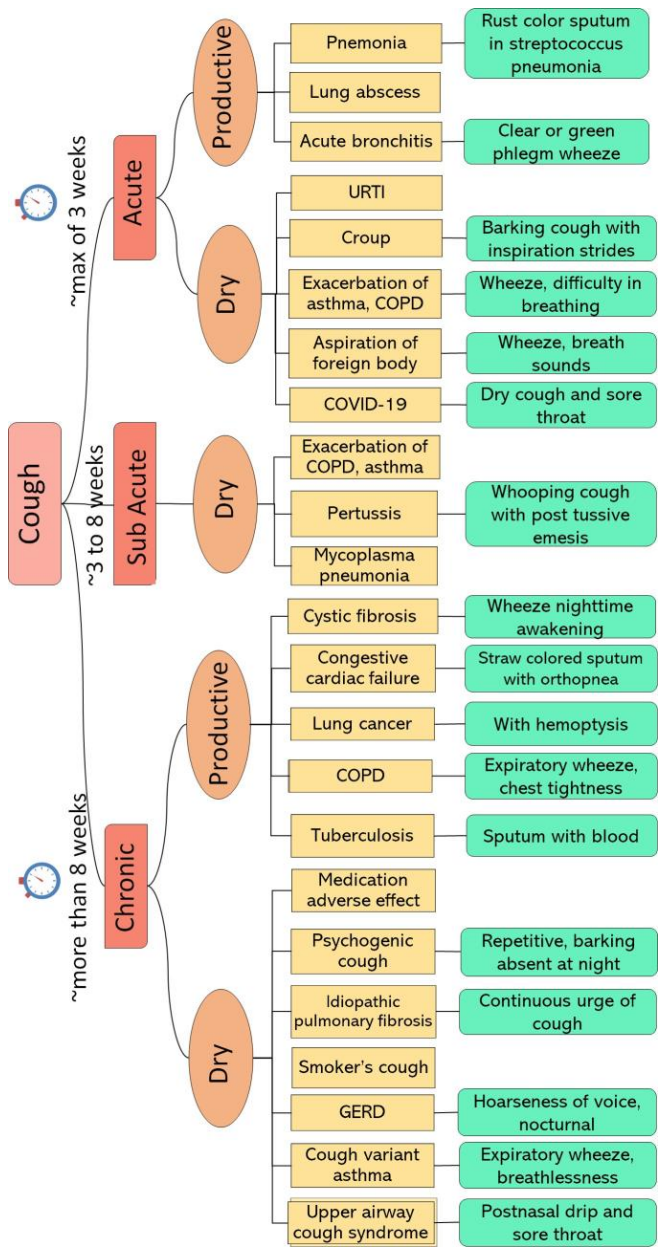

Fig. 6: Classification of cough according to the duration.

*C. Chronic Cough:*

Chronic cough lasts more than eight weeks. The causes of chronic cough include clinically severe diseases like lung cancer and less severe clinical causes like upper airway cough syndrome (UACS), cough variant asthma (CVA), bronchitis, smoker's cough, idiopathic pulmonary fibrosis, cough due to medications, habitual or psychogenic cough, gastroesophageal reflux disease (GERD), and obstructive sleep apnea [57]. A study done by the American College of Chest Physicians (ACCP) highlighted that upper airway cough syndrome, cough variant asthma, and GERD are the most common causes of chronic cough [116]. Physicians need to rule out the triad of these diseases in patients complaining of chronic cough [116]. CNN based imaging and cough analysis methods can assist the physicians for timely and accurate root causes of chronic cough.

- *Asthma:* It is a chronic hypersensitivity to allergens, dust, animal dander, and exercise. Its onset is mainly at an early age. The cough variant asthma patient presents with dry cough accompanied by breathlessness and a characteristic whistling sound called wheeze, which occurs in expiration. The symptoms occur characteristically at night. It is also associated with other atopic symptoms like eczema, allergic rhinitis, and food allergies. It is diagnosed by spirometry on which the FEV1/FVC ratio is less than 70%. Non-asthmatic eosinophilic bronchitis (NAEB) presents with similar clinical symptoms, but the patient has a productive cough with sputum eosinophili.

- *Gastroesophageal Reflux Disease:* The most common non-respiratory cause of chronic cough is gastroesophageal reflux disease (GERD). It can cause cough due to regurgitation and aspiration of gastric secretions resulting in irritation of vocal cord. The cough is non-productive, nocturnal, and accompanied by hoarseness of voice. 24-hour esophageal-PH monitoring can have high sensitivity and specificity in diagnosing GERD [65].

- *Upper Airway Cough Syndrome:* UACS presents with a dry cough, itching, sensation of dripping in the throat, sore throat, nasal stuffiness, nasal blockade, and rhinitis. It is a clinical diagnosis and in patients with atypical symptoms, diagnosis is made with an improvement of symptoms after prescribing first-generation oral antihistamines [77].

- *Chronic Obstructive Pulmonary Disease:* Smoking is a lung irritant that can cause a cascade of inflammatory reaction in the lung parenchyma. Heavy chronic smokers can develop a peculiar smoker's cough. It is chronic in onset and is continuous. Smoking and other occupational exposures can result in COPD. It includes both chronic bronchitis and emphysema. It presents with a productive cough containing small to moderate amount of mucoid phlegm, breathless on exertion, chest tightness, and expiratory wheeze. It is diagnosed by an in-office spirometry test. As per GOLD guidelines, an FEV1/FVC ratio of less than 0.7 after bronchodilator confirms the diagnosis of COPD [117]. Hence, there can be a chance of misdiagnosis of COPD, therefore, some additional AI-based methods are crucial for an expedited and accurate diagnosis to aid the state-of-the-art GOLD standards.

- *Tuberculosis:* One of the important diseases associated with the chronic cough is tuberculosis (TB). It has symptoms such as productive cough, weight loss, and low-grade fever for more than two weeks. State-of-the-art radiological investigations are carried out in the patients, along with sputum smear microscopy and culture. However, TB patients have very distinct cough types and by exploiting the latent features for the training of AI-based cough detection models, the diagnosis and treatment of this disease can be significantly improved.

- *Congestive cardiac failure:* It can cause cough due to pulmonary congestion and edema. It is associated with

11dyspnea, orthopnea, swelling in the legs, and fatigue. The cough in congestive heart failure has characteristic pink straw-colored sputum. It is diagnosed by cardiac evaluation and imaging such as echocardiography. By leveraging the expertise of health professionals and curated labeled databases related to imaging and cough, detection of congestive heart failure can be performed in significantly less time.

- *Idiopathic pulmonary fibrosis:* It is a multi-factorial disease, that can be caused due to the occupational exposure, smoking, or use of certain drugs. Pulmonary manifestations include cough, breathlessness, and hemoptysis. The cough is non-productive, dry, and continuous. Patients have a continuous urge to cough that is not alleviated by coughing. Patients may seldom have a small amount of clear sputum likely due to traction of bronchi [118]. An advanced imaging study like HRCT, lung biopsy, respiratory studies are carried out to confirm the diagnosis [102]. Hence, it is possible to focus on the AI-based technology for general image preprocessing procedures applied in chest radiography for disease diagnosis along with ML/DL-based cough detection methods.

- *Psychogenic cough:* Psychogenic cough is non-productive, barking and honking in nature, and is absent at night. As per the ACCP guidelines, the diagnosis of somatic cough should not be made only on clinical characteristics of the cough. Therefore, there is a dire need to exploit and enhance the existing databases for such diseases and train the ML models for better detection based on the latent cough features.

- *Cystic fibrosis:* Patients suffering from cystic fibrosis have productive cough (that has thick and dense sputum), wheezing, chest tightness, night-time awakening, and gastrointestinal symptoms of malabsorption. Newborn screening is done for cystic fibrosis, followed by a sweat chloride test to confirm the diagnosis.

- *Lung cancer:* Cough is mainly present in centrally located lung cancer like small cell and squamous cell carcinoma [119]. Productive cough with large amounts of mucoid sputum is characteristic of mucinous adeno carcinoma.

- *Cough caused by medications:* Finally another cause of chronic cough is different medications that can cause cough. When the cough is caused by angiotensin-converting enzyme (ACE) inhibitors or sitagliptin, the medication is discontinued for four weeks. If the cough improves within one to four weeks, the likely cause of cough is medication [65].

Given the distinct characteristic, symptoms, and existing medical diagnosis for the respiratory diseases mentioned above, and with limited provided resources in terms of medical professionals and testing equipment, there is a crucial need to leverage the development in Artificial Intelligence domain. Despite being in its infancy, the concept of using CAD and AI techniques is developing very fast. Machine learning and deep learning methods can be successfully used as tools to proactively aid healthcare professionals for better, fast, and cost-effective detection and diagnosis. In the subsequent sections, we have discussed the ML and DL models trained using cough samples that have been used in literature for the detection and diagnosis of the many above-mentioned respiratory conditions.

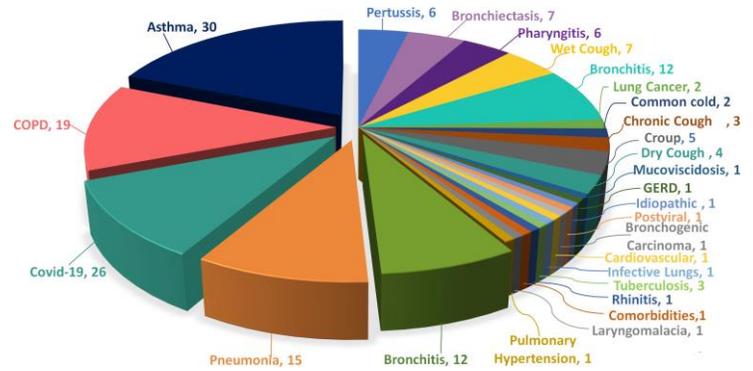

Fig. 7: Pulmonary diseases and other medical conditions associated with collected data in different studies

## IV. Data Acquisition for the Training of AI-based Cough Detection/Diagnosis Framework

The training of an accurate, reliable, and comprehensive cough detection/diagnosis model requires the collection of representative and relevant data. Data collection step is the first and foremost step in the ML/DL classifier development process, as shown in Fig. 5. This section provides the details about data collection devices, the characteristics of patients that contribute towards the dataset creation, and the customised apps/websites developed for the cough data acquisition.

### A. Data Collection

For the development of an effective ML/DL model, the training data must be representative, properly tagged, immune to the noise, and should include certain characteristics. There are mainly three approaches for data acquisition in the literature: data discovery, data augmentation, and data generation [120]. Data discovery phase includes searching for the new relevant datasets and it is recommended to make the datasets available online. Data augmentation complements the data discovery phase where the discovered datasets are not sufficient and require additional data from the external sources. Whereas the data generation is required when the external dataset are scarce or unavailable, however, there is a possibility of crowdsourcing and creating synthetic dataset.

Data collection comprises three major steps: *data acquisition*, *data labeling, and improvement of the existing data*. Data collection in the healthcare domain is a major bottleneck as it requires the continuous assistance from the trained medical professionals and consent from the patients to collect the data while they are undergoing some medical ailment. In the literature, cough datasets are collected from an extensive range of patients suffering from diseases such as pertussis, pneumonia, bronchitis, asthma, COPD, and COVID-19, to name a few. Fig. 7 enlists all the diseases which contribute to the data acquisition process in the articles that are included in this survey. In most of the studies, for the training of AI-based detection and diagnosis models, the cough acoustic data is collected from a variety of different diseases, of COVID-19 patients with the help of ML/DL frameworks. This is because cough is noticed as one of the eminent features in the diagnosis of the underlying disease and several studies have shown promising disease predictions with high accuracy [32], [39], [40].

TABLE V. A LIST OF COUGH DATA ACQUISITION DEVICES AND THEIR ATTRIBUTES

| Sensing Devices | Ease of use | Expensive | External help needed | Noise sensitivity | LOS limitation | Energy efficient |
|---|---|---|---|---|---|---|
| Smartphone | High | Medium | Low | Low | Medium | Low |
| Cough recording robot | Medium | High | High | Medium | Low | Medium |
| Piezoelectric Sensor | Medium | Low | Medium | Low | High | High |
| AioCare Spirometer | Low | Medium | Medium | Low | High | High |
| Hull Automatic Cough Counter | Medium | Medium | High | High | Medium | High |
| Microphone | Medium | Low | Medium | High | Medium | High |
| MP3 recorder | High | Low | Low | Medium | Medium | High |
| Portable voice recorder and microphone | High | Medium | Low | Medium | Medium | High |
| Smartphone customized App | High | Medium | Low | Medium | High | Low |
| Smartwatch | High | Medium | Medium | High | High | Low |
| Computerized Data Acquisition System | Low | Medium | High | Medium | Medium | High |
| Stethoscopes | Low | Medium | High | High | High | High |
| Public Telephone Hotline | Low | High | High | High | Low | Low |
| VitaloJAK wearable microphone | Medium | Medium | Low | Medium | High | High |
| Online platform | High | Low | Low | High | High | High |
| Microwave Doppler sensor | Low | Medium | High | High | High | Medium |
| Cough detection camera | Low | High | High | High | Low | Medium |

### B. Dataset Demographics

Cough dataset are collected from the patients all around the world, including Bangladesh [129], India [40], South Africa [43], US [130], China [131], [132], and Brazil [43] etc. The studies consider the fact that training of the ML models requires the dataset to be truly representative and inclusive. The database is generally collected from the patients in three environmental settings, i.e., hospitals, lab setting, and normal routine with varying levels of noise environment. Therefore, the data contain cough of all age groups ranging from infants of few months old to elderly patients of more than 60 years old [133]. The databases consist of three single classes; voluntary coughs, artifact and speech, and simulating the real working environment of system. However, the focus of most of the included studies in this survey involve children to teenage patients while some studies solely collected cough dataset from the elderly patients. Immense variation in the dataset size is observed among studies ranging from 4 patients [134], [135] to the crowdsourced data of 7000 individuals [39], and in one case more than 20,000 samples [136], more details about sample size is reported in Table VII-IX. For the evaluation of automated ML detection/diagnosis algorithms, some studies also collected the cough samples from healthy volunteers including both males and females [26], [137]–[141]. The reason for the inclusion of normal/healthy cough is to evaluate the ML model if it is trained enough to classify normal cough from the unhealthy cough produced by the patients. On one hand, the study that has the smaller datasets exploits the domain knowledge and transfer-learning to achieve the meaningful results [32]. Whereas, on the other hand, larger datasets can be useful for the training of deep learning models that can generate the features automatically, making feature engineering task easier.

### C. Open Source Databases

Despite of the escalated trend of using cough signals as a diagnostic tool for the respiratory conditions using AI techniques, the number of open source and validated cough dataset is limited. The outbreak of SARS-CoV-2 pandemic has put a tremendous strain on the healthcare systems, researcher from interdisciplinary domains got motivated to contribute by proposing efficient and cost-effective solutions. To help the community for developing the robust ML models, enormous amount of COVID-19 related data is required. Therefore, the researchers are motivated to collect the cough data and make it public to help the community in developing rigorous classifiers for the timely and expedited disease diagnosis. In this subsection, we list the open-access cough dataset:

13i. The COUGHVID dataset is the largest expert-labeled cough datasets in existence, comprised of more than 20,000 crowdsourced cough recordings, which includes a variety of demographic diversity [136]. The team collected the dataset, filtered it, and then labeled more than 2,000 recordings with the help of the expert pulmonologists. The database can be exploited for a numerous cough acoustic classification tasks.

ii. A public database Corswara comprised of approximately 1000 samples of respiratory sounds such as ,cough, breath, and voice [142]. By recording nine different sounds from each patient, this cough repository helps to extract the multi-dimensional spectral and temporal features. Its aim is to complement the PCR-based COVID-19 diagnosis methods. The sound samples are collected via worldwide crowdsourcing using a website application. The database is still in progress with the objective of better detection and quantification of the disease bio-markers with the help of sound acoustics.

iii. Cohen-McFarlane and team introduced a database named as NoCoCoDa, where the authors collected the cough events through the public media interviews with COVID-19 patients [143]. The data repository has 73 cough events and it is created by the manual segmentation, extraction, and annotation of data. They also investigated the severe cases of COVID-19, where the cough can be productive, as mostly the patients have dry cough. It is worth mentioning that this database is not public, however, it can be available on request.

iv. Another open source yet limited cough repository is developed by the independent AI researchers through website known as Virufy [144]. The dataset consists of only 16 patients (10 male and 6 female) with 7 COVID-19 PCR test positives while the remaining 9 are negative test patients. The biomarkers of the collected dataset are temperature, cough, shortness of breath, and glucose level.

In addition to the aforementioned public cough databases, Shuja et al. [145] provided extensive details about the open-access medical images, textual, and speech datasets for COVID-19. To ease the access of the scholarly articles, Allen Institute for AI created the COVID-19 Open Research Dataset (CORD-19) [146]. Despite of all the success in the data collection, it is evident that further efforts are still required to make the data collection process more systematic and accessible for the training of robust models. In addition, for the results to be reproducible the dataset should be public, that can assist in the development of vigorous ML/DL algorithms for the accurate disease diagnosis.

*D. Data Collecting Devices and Sensor placement*

Smart healthcare sensors with integrated circuit, optimal computational capability, and prolonged battery life are essentially required to acquire the data for the accurate AI-based predictions. A variety of data acquisition devices are used in the literature ranging from simple voice recorders [147] to a high-tech customized robot [148] for the cough collection and the associated disease diagnosis. These sensors have the ability to sense different environmental parameters and detect any changes within the proximity of the patient. We review some of the most commonly used cough data collection devices in Table V and investigate their efficacy in terms of cost, ease of use, noise sensitivity, energy efficiency, and Line-of-Sight (LOS) limit. Each wearable sensor or portable device has its own certain strengths and shortcomings. For instance, MP3 recorders are generally inexpensive, power efficient, and can tolerate high sound pressure levels, but are sensitive to high frequency and not immune to background noise [147].

Contrary to that, smartphones have the ability to capture continuous data, provide the dynamics of the real-life patterns, immunity to the background noise, readily available, user-friendly, and patient can place them within the close proximity. However, when the Signal to Noise Ratio (SNR) is low, the smartphone-based systems deactivate. Other challenges include the limited battery life and compatibility issues [149]. Cough collection using microphones and voice recorders is one of the most common practices in the literature [22], [123], [131], [132], [134], [137], [150]–[161]. The microphone can be placed near to the patient's chest for several hours [151]. In [123] to suppress the surrounding noise while placing the microphone on the chest, the authors covered the microphone with a plastic foam membrane. The microphone can be attached to the shirt collar [159], [162] in home/office setting or lapel [156], [157], [163] for 24 hours while performing the daily routine tasks. Due to the technological advancements and ease of access, smartphones have become ubiquitous. Therefore, the researchers have also exploited the smartphone data recording feature to collect the cough data [29], [30], [32], [121], [122], [127], [130], [164]–[172]. Smartphones have noise robustness for up to -15 dB and can easily be placed in the patient's pocket or the handbag for the data monitoring. However, as discussed earlier, the main challenge of using the smartphones for the data collection is limited battery life, the authors in [165] devise real-time cough detection algorithm that is power-efficient. Smartphone-based cough detection frameworks have the potential to accurately detect cough from the passively collected patient data and reduce the load of labeling the coughs manually. However, to make it easier for the patients to monitor their respiratory conditions and report about any medical anomaly accurately to the physicians and clinicians, further research is required in this domain.

Using microphone is also a common practice to acquire cough data either in hospital, lab or normal settings [135], [160], [173]–[177]. A high fidelity computerized data acquisition system which is comprised of low noise microphone, laptop, professional quality pre-amplifier, and A/D converter unit, is used for the sound signal acquisition in [178]–[181]. Other data collecting sensors include the FDA approved cough monitoring device VitaloJAK [124], smartwatch [170], [182], piezo sensor [138], public telephone hotline [183], hull automatic cough counter (HACC) [184], cough recording robot [148], microwave doppler sensor [141], AioCare spirometer [75], YouTube [185], cough detection camera [41], and stethoscopes [186].

Regardless of the devices used, the duration of the collected data varies from a few hours to a couple of months. The authors in [180] acquired the data by attaching the microphone near trachea for 4-6 hours while perna et al. [176] collected the audio cough sounds using a remote microphone for 90 days. Placement of the data acquisition devices and their distance to the patients are also a crucial consideration aspects for the data collection with minimal background noise. In [155] the microphone is attached to either the thorax of subjects or trachea while sitting or standing in the presence of ambient noise. The distance between the microphone and patients varied from 40 cm to 70 cm due to the inconsistent patient movements in the hospital [174], [175], [178], [181], [187]. However, in some circumstances, wearable sensors are not suitable for long-term use because of their weight and the requirement of keeping the sensor in close proximity to the patients, that hinders the convenient and continuous real-time data acquisition. By considering all these aspects, advanced and specialised data collecting devices need to be designed to ensure the convenience of patients and that can collect the data with minimal background noise for the accurate classification.

### E. Customized Developed Apps

The data-driven AI medical frameworks have shown great potential to become a powerful candidate to design, optimize, and adapt the state-of-the-art disease detection and diagnosis practices. These algorithms lead to a new paradigm with an ability to improve the healthcare and making sickcare treatments patient-specific. Nevertheless, machine learning models and neural networks, for the classification and pattern recognition purpose require sufficient amount of training data. Recently, the idea to design and develop accessible, easy to use, cost-effective, end-to-end customized applications and making their availability at patient's disposal, is gaining significant research interest. Smartphones being an ubiquitous device can be a prospectively effective source to capture the behaviors and dynamics of the people's real-life patterns. Smartphones allow researchers to collect active data in terms of self-reporting and passive data by monitoring the incidental or unconventional continuous contextual data. These distinctive features enable the opportunity to run the ML approaches within the smartphone devices. By downloading such specialized applications the subjects/patients are able to monitor the biomarkers and provide data, thus helping in building a corpus of the acoustic data. This becomes the motivation for the researchers working towards anytime, anywhere apps for the cough collection, detection or diagnosis. Fig. 8 summarizes the customized apps/websites for the cough data acquisition and the preliminary disease diagnosis. This subsection provides the details of these applications along with the relevant studies.

- *Mobicough*: A cost-effective mobile application is developed for the real-time detection and monitoring of the cough in [188]. For the data collection, an audio logging program is developed for the smartphones. Patients were asked to wear the earphone speakers while performing their routine works during 3-6 hours of the day. After the data collection, it further underwent four stages: *data pre-processing, segmentation, feature extraction, and classification for prediction*. Features such as Mel Frequency Cepstral Coefficients (MFCC), Zero Crossing Rate (ZCR), and entropy are then extracted and, later, based on these features, Gaussian Mixture Model (GMM) is trained for the modeling of cough sounds. For the modeling of background sounds such as noise and speech, Universal Background model (UBM) is employed. Results showed that the system is capable of cough detection using smartphone app, however, the authors have not compared the performance of the developed app with any commercially available cough detectors.

- *Healthmode*: A smartphone customized application is designed for the continuous cough data collection and detection analysis [189]. To classify the spectrograms and for the recognition of cough in the recordings, the authors used Convolutional Neural Network (CNN). The data is collected by using the phone's internal microphone. After the recording, the cough data is sent to the secure cloud server for further analysis. Finally, the performance of the devised technique is compared with the six commercially available cough monitors, namely, Lifeshirt, Vitalo-Jak, Leicester Cough Monitor (LCM), Cough COUNT, HACC, and LR100 Cough Monitor. The results in terms of sensitivity (90%), specificity (99.5%), and inter-rater agreement have showed comparable performance with the existing commercial solutions. In addition, while developing the app, the authors ensured that the privacy of the data remains intact, a feature that is of paramount importance.

- *COVID-19 sounds*: Researchers in [39] has developed a cross-platform application for the collection of crowd-sourced database of more than 10,000 cough samples from around 7,000 unique users till date. Out of this large dataset, 235 subjects are diagnosed with COVID-19, roughly comprised of 3.36% of the entire dataset. The authors assert that it is possible to classify COVID-19 patients form the healthy users, and the cough of COVID-19 patients from that of asthma cough with appreciable performance, even by leveraging simple ML models such as Logistic regression model (LR) and Support Vector Machine (SVM) trained with the handcrafted features that are extracted using transfer learning. The proposed model is evaluated by using a subset of the datasets (141 cough and breathing samples). The models achieved an AUC (Area Under the Curve) of above 80% for all the tasks. Another distinctive feature of this custom made mobile application is that it reminds the users to provide the samples after every couple of days. However, this dataset is not yet open source, therefore, it cannot be used to train the ML and DL models for the sake of reproducibility of the results.

- *AI4COVID-19*: An application to capture the coughs of the patients suffering from bronchitis, pertussis, COVID-19 was developed in [32]. The authors investigated the distinctness of pathomorphological alterations in the respiratory system caused by SARS-CoV-2 infection when compared to the other respiratory conditions. In addition to collecting the cough data, this application proposed the diagnose of COVID-19 based on cough data. For the detection of coughs from other sounds, the authors trained a CNN model. While the diagnosis architecture leveraged a multi-pronged mediator centered risk-averse AI framework, showing an accuracy of more than 90% for the disease diagnosis. Transfer learning is leveraged to deal with the scarcity of the COVID-19 cough training data. This diagnosis framework is comprised of three independent classifiers, namely, Deep Transfer Learning-based Multi Class classifier (DTL-MC), Classical Machine Learning-based Multi Class classifier (CML-MC), and Deep Transfer Learning-based Binary Class classifier (DTL-BC). To create an autonomous model, if the output of any two classifiers mismatches then the result turned out to be inconclusive. Thus, the app shows promising results for the preliminary diagnosis of SARS-CoV-2 and has the potential to assist the medical practitioners in timely treatment of the patients.

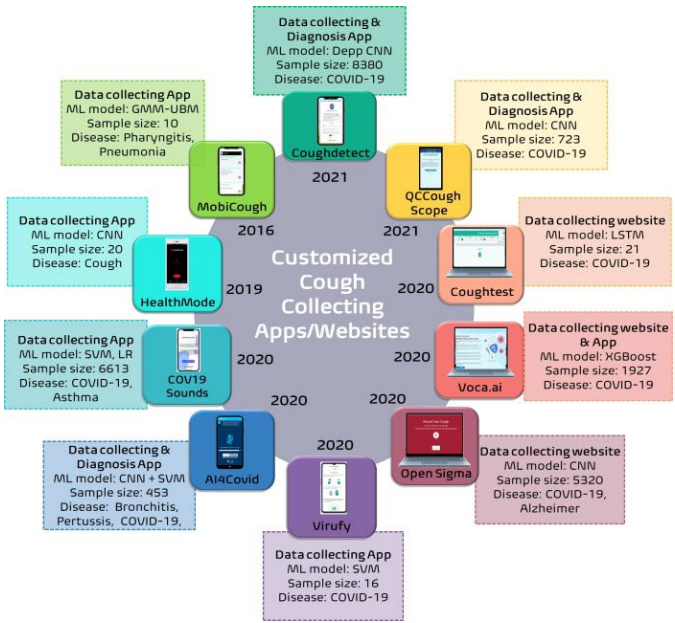

Fig. 8: Customized cough detection and disease diagnosing applications.

- *Virufy*: An application and study developed by the researchers from 25 countries with an aim to help the low-income countries fighting against COVID-19 pandemic outbreak. The app encourages the people from eight countries to record their cough sounds using smartphones and help to train a robust AI algorithm for better cough pattern recognition. By using this Virufy dataset together with the other publicly available datasets, i.e., Coswara and Coughvid, the team in [190] developed a deep neural network to show the applicability of the crowd-sourced datasets for the detection of COVID-19 based on the cough data only. Commonly used audio features, i.e., MFCC and mel-frequency spectrograms, are exploited for the model training.
  The trained AI-based method accurately predicted the COVID-19 infection with an ROC-AUC of 77.1%. Another study by the same research team is also performed to develop an ensemble learning architecture based on a residual-based neural network architecture (ResNet-50) CNN [191]. The model is trained using three different types of sample data, i.e, COVID-19 positive, symptomatic COVID-19 negative, and asymptomatic. The model achieved remarkable performance in terms of AUC (99%) for COVID-19 samples. For the future work, these authors aim to improve the algorithm in a way that it could diagnose the disease based on the cough data with unavoidable background hospital noise.

- *Coughtest*: In [43] the authors collected coughs from the patients who underwent SARS-CoV-2 virus test, using the application Coughtest [192]. Before recording the cough, the subject has to answer the questionnaire prompted on the website regarding age, gender, country of residence, smoker or non-smoker, and some specific questions, for instance, if a person has COPD, lungs cancer, cystic fibrosis, COVID-19, or has any symptoms. Collected data from this app and Coswara crowdsourced data result in a combined data from the six continents. After the normalization of data, features are extracted to train several classifiers; LR, SVM, Multilayer Perceptron (MLP), Long Short Term Memory (LSTM), CNN, and a Resnet50. LSTM demonstrated the best performance in discriminating the COVID-19 positive coughs from COVID-19 negative cough as well as from the healthy cough with an AUC of 0.94 based on 13 sequential forward search (SFS) best features. The researchers are doing continued efforts to make the models more robust and implement the classifiers on the smartphone platform.

- *Coughdetect*: A web-based full-stack automatic processing screening tool is developed for the detection and diagnosis of COVID-19 cough samples [193]. The subjects having symptoms such as cough, fever, shortness of breath etc. recorded their cough and sent to the server to have a preliminary diagnosis about the disease. The cough is then processed for the analysis and prediction. Finally, the user of the app received an asynchronous message through a secured connection informing if he/she has COVID-19 or not, and whether they should consult a physician. The authors extracted three sonographs namely MFCC, Melscaled spectrogram (Melspec), and Linear Predictive Coding Spectrum coefficients (LPCS) for the model training. In addition to cough, the authors used other biomarkers to predict the severity level of the disease. Based on the cycle threshold (Ct) from the qRT-PCR test or lymphocyte counts, the data is labeled to predict the extent of the infection: borderline positive, standard positive, or high positive. The data labeled based on qRT-PCR COVID-19 diagnosis results showed an encouraging AUC of 98.80% for the recognition of the three disease severity levels. The proposed primary test can mitigate the cost of the COVID-19 diagnostic test and can make the processing faster.

- *Opensigma*: A project by MIT researchers, aimed to diagnose COVID-19 by training the CNN model based on the cough data collected through website *opensigma.mit.edu*, which contains one of the largest COVID-19 cough datasets, comprised of 5,320 subjects. Transfer learning was applied for AI speech processing framework for the feature extraction. The CNN-based framework is trained with the data collected from the COVID-19 and Alzheimer's patients. The architecture is consisted of one poisson biomarker layer and 3 pre-trained ResNet50s in parallel. The trained model showed promising results in diagnosing COVID-19 with a sensitivity of 98.5% and a specificity of 94.2%.

- *Voca.ai*: To enable the swift diagnosis of COVID-19, Voca and researchers from Carnegie Mellon University worked on a project Voca.ai, which is an application and the website that asks the volunteers for the cough recordings [194]. The authors trained the XGBoost model using the crowdsourcing data with 1927 samples uttering vowels/sounds such as, "ah", "oh", and "eh". The authors investigated the use of the symbolic recurrence quantification by extracting MFCC features to detect COVID-19 in the cough signatures of the healthy and unhealthy people. The performance analysis revealed that the model has the capability of detecting the underlying dynamics in the vocal sounds and can effectively detect COVID-19 with an overall mean accuracy of 97%.

- *QUCoughScope*: Recently, an android application is developed by the researchers of University of Qatar to distinguish the COVID-19 patients from the other lung infections and normal people by collecting the cough and breathing acoustics. The data is comprised of the samples of 582 healthy individuals and 141 COVID-19 patients, out of which 87 are asymptomatic while the rest were sysmtomatic. The application asks for some symptoms-related questions and to record the audio signals, which are then sent to the server for the conversion of audio signal to spectrogram. To increase the sensitivity of the symptomatic COVID-19 patients and reduce the miss-classification, they deployed a cascaded pipeline comprised of three state-of-the-art novel deep CNN models. The overall sensitivity of the system reported by this study is 95.86%.

These customized cough detection frameworks have the ability to monitor the symptoms and passively collect the data from patients using smartphones or websites. This approach can reduce the load of labeling the coughs from the medical healthcare personnel. Thus, the sophisticated AI frameworks have the potential to provide preliminary diagnosis based on the collected cough signatures and can assist the medical practitioners in better decision making.

*F. Data Pre-processing*

The completion of the data acquisition step leads to the next crucial step within the data analytics pipeline, i.e., data annotation and pre-processing. This step has paramount importance as it is required for the refinement of the cough samples and for designing a high quality robust prediction model, as shown in the Fig. 5. Data pre-processing helps to boost the consistency and accuracy in the collected data, thus providing ease in the data interpretability [195]. In general, there are four main steps in data pre-processing: 1) data cleansing, 2) data integration, 3) data reduction, and 4) data transformation [196]. However, in the literature researchers adopt some steps while skipping the others, depending upon the data they collect from the sensing devices. During the pre-processing, the acoustic data is segmented into frames by isolating the part of the captured signal from the silence. It is a step that delineates the region of interest, i.e., cough events. For this step various tools and speech analysis software packages are used, for instance, Audacity, PRAAT [197], WEKA [170], Wavesurfer [198], and MATLAB Signal processing tool-box [184]. These tools are used to remove the silent portion at the beginning and end of the captured recordings, and for the rescaling and down-sampling. PRAAT is also widely utilized for the data annotation [130]–[132], [165], [188], [197] In order to successfully perform this step, the acoustic data is segmented into frames, RMS energy for the acquired frame is calculated and compared with a predetermined thresholds. In case the frames have low energy, it is assumed to be silence or ambient sound and, therefore, are discarded. High energy windows are selected for further processing and feature extraction [138]. To further remove the noise and dispose of the minor artifacts, filtration is applied. Finally, the data is normalized within a given range. The above mentioned pre-processing steps are considered very significant as they help in obtaining a well refined and transformed cough signals that are free from any ambient noise or silence. These pre-processed frames can then be used for the feature extraction and data analysis. Hence, this step is essential to improve the accuracy and performance of the prediction models.

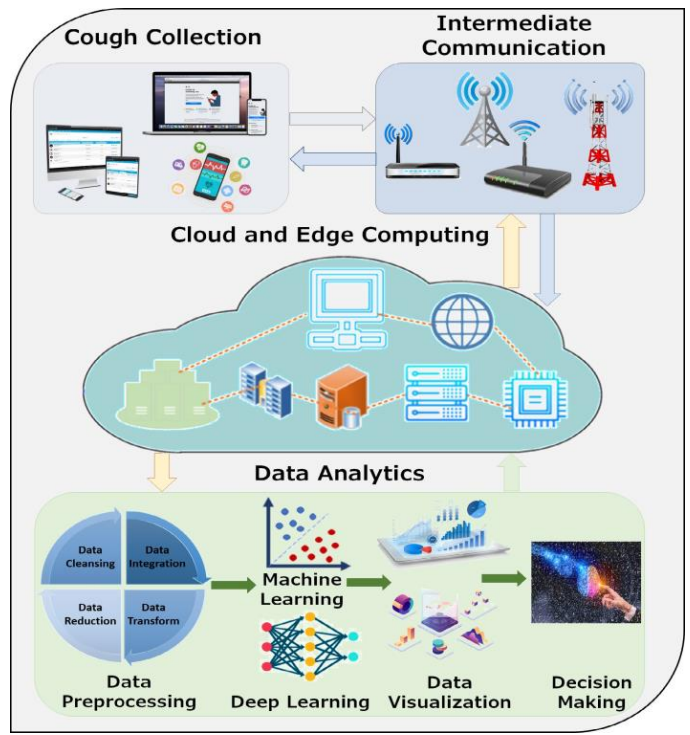

Fig. 9: A comprehensive smart healthcare framework involving Artificial Intelligence and Cloud Computing.

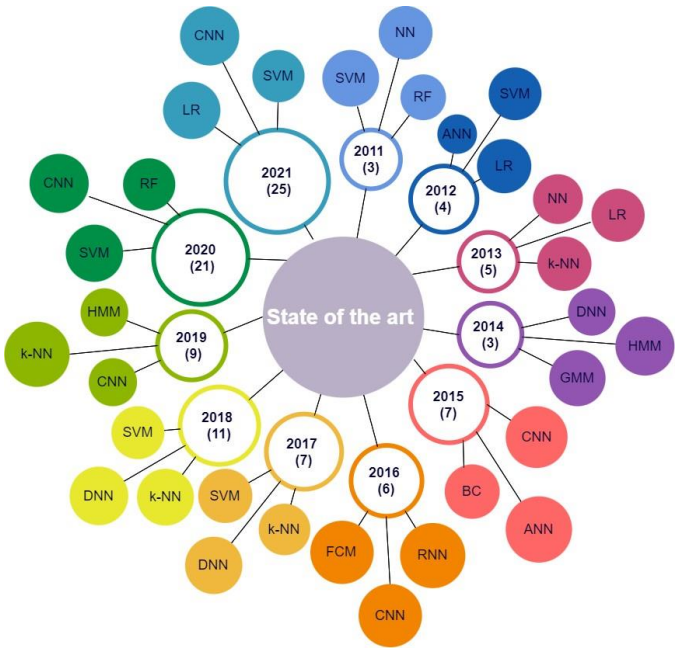

Fig. 10: Data-driven cough-based AI literature over the years.

### G. Smart AI-Based Healthcare Solutions involving Edge/Cloud Computing

It can be noted that after the successful collection of the data, it is further analyzed, for example, for correct labeling and annotation. If this is handled manually because of the restricted resources, then it leads to high latency and prevention of real-time accessibility of the smart solutions developed for accurate detection and disease diagnosis. Therefore, an end-to-end interconnected, reliable, accurate, and low latency solution is required for the assistance of existing medical practices. Progress in concurrently-advancing technologies such as cloud/edge computing and mobile wireless communication can be incorporated in enhancing the diagnostic accuracy of healthcare-based AI classifiers by providing the right type of data from the patients in a timely manner [199]. An end-to-end comprehensive health monitoring framework is shown in Fig. 9. It comprises of data acquisition devices that, as a first step, collect data from the patients. After the successful collection, depending upon the communication range and medium, the data is then transmitted through either wired or wireless connections to the cloud for its further processing including continuous detection and classification using AI techniques [200].

In some cases, to perform extensive processing and classification tasks, the collected data is sent to the ML backend core to employ the AI-based data-driven algorithms for an in-depth analysis and decision making [201], [202]. The new paradigms of edge/cloud computing provide innovative solutions by bringing the resources closer to the patients for symptoms monitoring as well as to help the medical professionals by providing assistance in the form of low latency and energy-efficient solutions [203].

The aim of devising such all-inclusive smart healthcare systems is the automation of collection of patient's vital data. Additionally, it is also made sure that the data is delivered using a secure, fast, and reliable medium to cloud for its further processing as well as storing, if required. Several advantages of such extensively connected and distributive framework involving the edge/cloud computing include: 1) automatic collection of real-time continuous representative data without draining the power resources of the data collecting devices, 2) elimination of manual data annotation to eradicate the issue of limited healthcare trained workers, and 3) provision of uninterrupted detection and diagnosis services with noticeable accuracy where the medical resource/facilities are scarce. Thus, such frameworks open up a paradigm to enable robust tele-medicine setup along with the facilitation of the medical experts using data-driven edge-enabled technologies.

## V. DETECTION AND DIAGNOSIS USING ARTIFICIAL INTELLIGENCE TECHNIQUES

Before the advent of artificial intelligence techniques, predictive modeling in healthcare is merely considered as a source of provision of limited automated tasks such as data cleaning and automated organization of the health data. However, sophisticated machine learning/deep learning tools have shown the ability to learn extremely complex relationships in performing some medical related decision. These intelligent data-driven medical models have the potential of pattern recognition and interpretation of the raw unstructured data. Thus, AI demonstrate encouraging results to facilitate the physicians for providing the support in decision making. These aspects have attracted a lot of attention and motivated the researchers to analyse the applicability of the ML models in the medical domain.

In the last decade, a significant amount of work has been done using conventional ML techniques and Neural Networks (NNs). Fig. 10 summarizes the number of the peer-reviewed published papers in each year over the last decade that developed AI models trained by using the cough signals. This figure also shows the three most commonly implemented ML/DL classifiers for the cough detection and disease diagnosis. The following section elucidates the state of the art AI techniques for cough-based disease detection/diagnosis. Furthermore, it discusses the classification of various ML/DL algorithms implemented in the literature, as illustrated in Fig. 11. Finally, it also provides the details of the types of features used to train the models and the performance metrics required for the validation of the AI-based frameworks.

### A. Types of Features

Cough is associated to over 25 respiratory syndromes and their underlying conditions. Different coughs caused by distinct pathomorphological alternations manifest distinct latent features [204], [205]. These features with discriminating characteristics can be extracted by employing the pertinent signal processing and statistical cough sounds transformations, to be used for the training of a sophisticated AI engine. Fig. 11 shows several types of the distinct cough features that can be leveraged for the development of ML/DL models.

These features can be broadly categorized as time domain and frequency domain features.

Some of the time-domain features that can be extracted are absolute mean, absolute median, standard deviation, skewness, kurtosis, and zero-crossing rate; and frequency-domain features such as spectral centroid, spectral roll-off, spectral variance, MFCC, and spectral chroma [170]. Spectral/Cepstral features which are based on the frequency and are obtained by converting the time domain signal using the frequency domain Fourier Transform. In [131] the authors exploit the fact that the cough signal is spread out in the entire frequency band, making the cough signal behaviour distinctive from the speech signal. Therefore, they used the spectral structure as filter banks of feature extraction methods for the cough detection from the other acoustic signals.

Cepstral features are also widely used to separate the cough features from other sounds like laughing, speech, and background noise. In [22], [28], [32], [159], [206] the authors exploit commonly using cepstral features such as MFCC, Linear Prediction Cepstral Coefficients (LPCC), and Gammatone Frequency Cepstral Coefficient (GFCC). MFCCs can be obtained by frequency transform of the log spectrum and these features take account for the non-linear response of the audio spectrum [207]. LPCCs are an extension of the linear prediction via autoregressive modeling in the cepstral domain [208]. The underlying differences among these features lie in the frequency representation scale and the feature dimensionality depends on the value of inner parameters values [167]. Apart from the aforementioned multidimentional features, there are some unidimentional features which are leveraged in [167], for instance, Spectral Standard Deviation (SpecSD), Spectral Skewness (SpecSkew), Spectral Kurtosis (SpecKurto), and Spectral Peak Entropy (SpecPeakEn), to train the k-NN model and demonstrated satisfactory performance.

To capture the long term dynamics of the cough sound signals, temporal features such as zero-crossing rate, energy of the signal and maximum amplitude can also be used for the model training [180]. In [186], the authors performed the quantification of noise in the events containing cough signatures by computing the noise features like spectral flatness, chirp group delaying, and harmonic to noise ratio are computed using the Voice Sauce toolkit. Other experiments based on both synthetic data and real data, demonstrate that ensembling of multiple frequency sub-bands has better performance than common feature extraction methods, such as MFCC and GFCC.

### B. Feature Extraction

Feature engineering is considered as one of the cornerstones for the development of ML/DL models. It is a process of extracting the most dominating and discriminating characteristics of an audio signal to obtain a suitable cough sound representation for the classification. To perform feature extraction, the cough waveform is modified to a relatively minimized datarate, for further processing and analysis. Features can be extracted from either time-domain signals or from its frequency-domain representation. Some of the commonly used feature sets extracted from the audio and speech signals are: MFCC, Linear Prediction Coefficients (LPC), LPCC, Line Spectral Frequencies (LSF), Discrete Wavelet Transform (DWT), and Perceptual Linear Prediction (PLP) [209]. As a general practice, for the generation of the features/sonographs, the authors leverage feature extraction libraries or sometimes exploit handcrafted features [138]. In [210] detailed description about the feature extraction libraries, suitability for various cases and their comparison is provided. Some of the frequently used libraries in the literature are: *Hidden Markov Toolkit (HTK) [151], [157], [158], [183], Librosa [32], [40], [206], Taros-DSP to extract MFCC features, jMusic to extract the spectral features [170], RASTAMAT Matlab Toolkit [131], OpenSMILE [211], and others include Meyda, Aubio, and LibXtract.*

Feature extraction libraries have certain distinct characteristics, i.e., Librosa helps to extract high-level and low-level features, however, it does not cluster the features. For the real time feature extraction, Meyda and LibXtract are more appropriate libraries, while for high level features and segmentation, Aubio is the suitable choice. Combination of LibXtract and Marsysas can also be used for the data annotation and visualization of the audio cough features [210].

### C. Categorization of ML/DL Algorithms

It can be noted that the psychological behavior of even same patient can be different in different environments while coughing, for instance, trying to resist the cough in the crowded environment or putting the elbow in front of the mouth, as a result, volume of the cough sound can be different. Therefore, the cough detection and disease diagnosis methods should be robust and highly sensitive to detect the cough acoustic data with minimal false alarm rate. In recent years, extensive research has been done in this domain to develop the automated AI-based screening tools for the timely and efficient detection/diagnosis of the respiratory conditions. Cough detection and diagnosis are essentially binary and multi-class classification problems, respectively, where the key features (shown in Fig. 11) are used as inputs to the machine learning model, to learn the complex behavior of the cough audio signatures for the respiratory disease diagnosis. In Fig. 11, we provide the categorization of the AI algorithms being implemented in the literature for the tasks of cough detection and disease diagnosis. The figure shows a range of different ML models 1) linear ML, 2) non-linear ML, 3) deep learning, 4) ensemble, and 5) other statistical methods that are leveraged for designing an ML-based disease diagnosis frameworks. In Table VII, we enlist the summary of the articles for cough detection; ML/DL algorithms, associated disease, the sample size being used for the model training and the performance evaluation. Table VIII presents the comparison of the AI models for the cough-based disease diagnosis (classification) using similar evaluation criteria, whereas, Table IX shows the summary of the articles that implemented ML models firstly for the cough signature detection from other similar acoustic signals and then implement the separate classification model based on the cough sounds for disease diagnosis. Insights regarding performance and computational complexity for each of these ML algorithms are out of the scope of this article, the interested readers can refer to the references given in the respective tables. In addition, the merits and demerits of these algorithms also cannot be fairly determined because of the difference in the underlying assumptions, sample size, and the approach taken to perform the experiments. For example, linear ML methods algorithms, such as Linear Regression, assume the training data to be of a specific functional form with a fixed size of parameters, whereas Non-linear ML algorithms, such as k-Nearest Neighbors and Decision Tree, are free to assume any functional form of the training data. Ensemble methods are of two types: Bagging and Boosting, which further have several variants, some of which have been listed in Fig. 11. Deep learning methods on the other hand, such as CNN and DNN, are universal learners. These algorithms are of significant value when a lot of training data is available, as they do not require manual feature engineering and are capable of latent feature extraction on their own. Some papers such as [133], [153], [159], [188] exploited hybrid or ensemble-based methods. Other statistical methods (also listed in Figure 11) are also used in earlier works on cough detection and diagnosis [30], [129], [150], [162], [188]. Among the most common combinations of feature set and ML model is MFCCs + SVM and the neural networks, which have the potential to model and achieve high accuracy. The performance of implementing different combinations of features and AI models such as MFCC+SVM, STFT+SVM and STFT+CNN are also combinely studied in [138], and the authors demonstrated that the CNN has the highest performance.

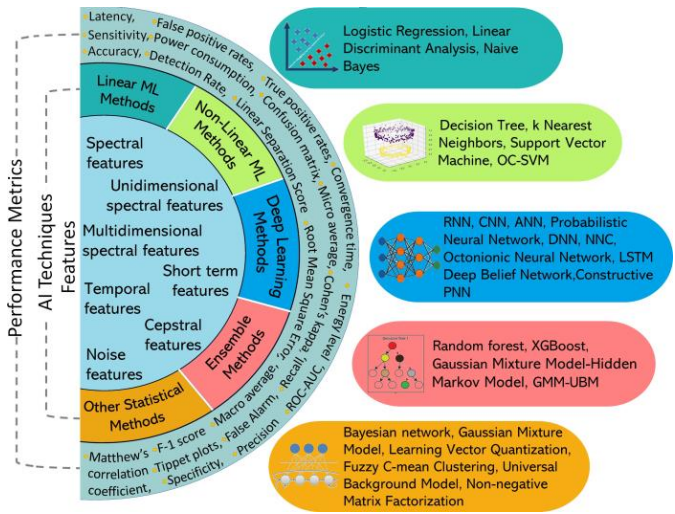

Fig. 11: Categorization of the AI techniques and performance evaluation parameters.

*D. Performance Metrics*

Once the ML/DL are implemented and trained, the next logical step is to evaluate the effectiveness of the model based on metrics and datasets. Based on the implementation differences and training data, different performance metrics are used for the evaluation of the ML/DL models. A list of metrics is also given in Fig. 11, among these the most widely used performance measures are accuracy, F1-score, sensitivity, and specificity [32], [75], [194], [213]. Cohen's kappa coefficient is another performance metric that overcomes the problem of overestimating the accuracy [194], [225]. We enlist the accuracy, sensitivity, and specificity in Table VII, VIII, and IX for the articles performing cough detection, cough-based disease diagnosis, and the articles that develop a detailed AI-based model for the cough detection along with the cough-based disease diagnosis, respectively.

In [122], [165], authors proposed an energy-efficient cough detection framework that consumes minimal power of the smartphone. For some cough detection models, the performance is compared with the commercially available cough counters such as LifeShirt, LCM, HACC, VitaloJak in terms of precision, recall, and accuracy in [130], [155], [165], [189], [226]. In addition, [170] evaluated the performance of the proposed algorithm based on the power and memory consumption and [227] proposed CoughNET, a CNN-LSTM processor, that consumes much less power and energy (290 mW and 2 mJ) compared to the other implemented methods for cough detection in the literature.

Another insightful performance measure is the confusion matrix, which is the correlation between the predictions of a model and the actual class labels of the data. Confusion matrix has the information of true positive, true negative, false positive, and false negative in itself and can be helpful to calculate the accuracy of the classification model. Hence, performance evaluation is an integral part of developing an AI-based framework. Also, the choice of metrics that are selected for the evaluation influences the performance of the machine learning models. It helps to select the ideal model that has the ability to correctly represent the data and depicts how accurately the selected model will work in the future, once incorporated in the real world healthcare applications.

## VI. DETECTION AND DIAGNOSIS USING NON-COUGH SOUNDS

Due to vocal cord dysfunction and respiratory issues, sounds other than cough are also produced by the vocal cords and blockage of the lungs airways. These sounds include noisy breathing, crackles, wheezing, stridor, pleural friction rub, and snoring. On one hand, these sounds can be misunderstood as cough making it difficult for a cough detector to separate cough from other sounds for further analysis. On the other hand, we can exploit these sounds along with the cough signatures for further research and better diagnosis of numerous diseases. In this section, we briefly discuss the literature that leverage non-cough sounds for the purpose of detection and diagnosis of different diseases.

- *Lung Sounds*: A systematic review discussed the details of the lung sounds classification/characteristics [15]. The authors also presented the machine learning techniques developed for the analysis of lung sound anomaly. Obstruction in lung airways may cause sounds like crackles, stridor, and wheezes. In [21], the authors performed a study for the recognition of wheezing sound in children using smartphones. They employed a two-phase algorithm, with signal performance analysis and SVM training. They achieved 71.4% sensitivity and 88.9% specificity. The researchers from Digital Health Lab Samsung performed a feasibility study for developing a mobile sensor framework based on wheezes with the characterization accuracy of 94.6% and recall of 74.62% [228].

TABLE VII. OVERVIEW OF RELATED WORK ON COUGH DETECTION ALGORITHMS

| Author | Best Performing AI technique | Disease Diagnosed | Sample size | Accuracy | Sensitivity | Specificity |
|---|---|---|---|---|---|---|
| Drugman et al. 2012 [155] | ANN | Cystic Fibrosis | 32 | – | 91% | 95% |
| Swarnkar et al. 2013 [180] | Neural Network | Infective lung disease, pulmonary oedema, COPD | 3 | 98% | 93.44% | 94.52% |
| Liu et al. 2014 [132] | DNN | Pneumonia, bronchial asthma, COPD | 20 | – | 90.1% | 88.6% |
| Sterling et al. 2014 [157] | HMM | Asthma | 29 | – | 85.7% | – |
| Drugman et al. 2014 [156] | ANN | Cough | 2338 samples | – | 91% | 95% |
| Amoh et al. 2015 [138] | CNN | Healthy cough | 14 | – | 95.1% | 99.5% |
| Wang et al. 2015 [131] | CNN | Pneumonia, bronchial asthma, COPD | 26 | 98.59% | 98.17 | – |
| Amrulloh et al 2015 [181] | ANN | Pneumonia, rhinopharyngitis, pulmonary hypertension, bronchiectasis, bronchitis | 14 | | 93% | 98% |
| Ferdousi et al. 2015 [129] | Bayesian Classifier | Cough | 1027 samples | 86.31% | 85.22% | 87.37% |
| Alvarez et al. 2016 [164] | k-NN | Cough | – | – | 94.17% | 92.16% |
| Amoh et al. 2016 [139] | CNN | Healthy cough | 14 | 87.6% | 82% | 93.2% |
| Pham et al. 2016 [188] | GMM-UBM | Pneumonia, pharyngitis | 10 | – | 91% | – |
| Amrulloh et al. 2016 [187] | Fuzzy c-mean clustering | Dry and productive cough | 39 | 76% | 77% | 75% |
| Rocha et al. 2017 [186] | LR | COPD, congestive heart failure | 59 | – | 92.3% | 84.7% |
| You et al. 2017 [159] | SVM | Pneumonia, bronchial asthma, COPD | 18 | 81.5% | 78.5% | 84.5% |
| Alvarez et al. 2018 [168] | k-NN | Cough | – | – | 92% | – |
| Alvarez et al. 2018 [121] | SVM | Asthma, COPD, bronchiectasis | 13 | – | 92.71% | 88.58% |
| Alvarez et al. 2018 [167] | k-NN | Asthma, COPD, bronchiectasis | 13 | 95.28 % | 88.42% | 96.8% |
| Barcelo et al. 2018 [165] | k-NN | Asthma, COPD, bronchiectasis | – | 95.07% | | |
| Barcelo et al. 2018 [122] | k-NN | Asthma, COPD, bronchiectasis | 13 | – | 88.94 | 98.64 |
| Klco et al. 2018 [123] | ONN | Asthma, COPD, bronchitis, lung-cancer, pneumonia | – | – | 96.8% | 98.4% |
| Kadambi et al. 2018 [124] | DNN | Chronic cough, COPD, asthma, lung cancer | 9 | 92.3% | 97.6% | 93.7% |
| Miranda et al. 2019 [74] | CNN | Cough | – | 91.2% | – | – |
| Teyhouee et al. 2019 [212] | HMM | Cough | – | 78% | 89% | 74% |
| Khomsay et al. 2019 [160] | DL | Productive cough | 8 | 96.88% | | |
| Kvapilova et al. 2019 [189] | CNN | Cough | 20 | – | 90 | 99.5 |
| Rahman et al. 2019 [170] | RF | Asthma, COPD | 131 | – | 94.1 | – |
| Barata et al. 2019 [213] | CNN | Cough | 43 | 90.9% | 91.7% | 90.1% |
| Vhaduri et al. 2020 [140] | RF | Cough | 25 | 96% | – | – |
| Drugman et al. 2020 [147] | GMM | Mucoviscidosis | – | – | 95.2% | – |
| Vatanparvar et al. 2020 [214] | GMM-UBM | Asthma, COPD | 131 | 84.89 | 72.17 | 97.61 |
| Chuma et al. 2020 [141] | CNN | Cough | 10 | 86.5% | – | – |
| Solinski et al. 2020 [75] | ANN | Asthma, COPD | – | 91% | 86% | 91% |
| Chen et al. 2021 [215] | SVM | Cough | 670 | 94.9% | 97.1% | – |
| Lee et al. 2021 [41] | CNN | Cough | 200 events | 96% | 90% | – |
| Xu et al. 2021 [216] | CNN & MobileNet | Asthma, COPD | 200 events | 94.9% | 91.2% | – |

- *Breath Sounds*: These sounds can provide significant information for the respiratory diseases such as flu, pneumonia, and bronchitis etc. In [229] an automatic breath sound detection system based on hybrid perceptual and cepstral feature set (PerCepD) is proposed. The authors trained SVM and Artificial Neural Network (ANN) models for the classification with high accuracy. In [20] an ML-based diagnosis mechanism is devised for pneumonia; a low-cost smartphone is used to collect breathing sound data. Features from various domains such as Teager energy operator-based, prosodic, spectral, cepstral, and delta-delta coefficients features are extracted. SVM and k-NN classifiers are trained to diagnose pneumonia patients with satisfactory accuracy. Healthy and COPD subjects are distinguished based on the respiratory sound analysis by employing machine learning techniques in [230]. The system comprises of offline and online units. In offline systems, initially, preprocessing and feature selection techniques are applied to train SVM, LR, k-Nearest

TABLE VIII. SUMMARY OF THE LITERATURE ON COUGH DIAGNOSIS ALGORITHMS

| Author | Best AI Tech | Disease Diagnose | Sample size | Accuracy | Sensitivity | Specificity |
|---|---|---|---|---|---|---|
| Swarnkar et al 2013 [179] | Logistic Regression | Asthma, pneumonia, bronchitis, rhino-pharyngitis | 46 | – | 72.7% | 79% |
| Parker et al. 2013 [185] | k-NN | Croup, pertussis | 47 | – | 75% | 100% |
| Abeyratne et al. 2013 [175] | Logistic Regression | Pneumonia | 91 | 94% | 93% | 90.5% |
| Koshaish et al. 2014 [135] | Logistic Regression | Pneumonia | 91 | – | 94% | 96% |
| Amrulloh et al. 2015 [26] | ANN | Asthma, pneumonia | 18 | 94.4% | 88.9% | 100% |
| Schroder et al. 2016 [183] | GMM | Dry and productive cough | 514 events | Dry cough: 33% <br> – | Dry cough: 91% <br> Prod cough: 6% | Dry cough: 15% <br> Prod cough: 49% |
| Sharan et al. 2017 [28] | SVM | Croup | 1364 | 91.21% | 88.37% | 91.59% |
| Windmon et al. 2018 [29] | RF | COPD, congestive heart failure | 36 | 78.5% | 82% | 75% |
| Sharan et al. 2018 [169] | SVM | Pneumonia, asthma/RAD, bronchiolitis, croup, URTI | 479 | 86.09% | 92.31% | 85.28% |
| Botha et el. 2018 [217] | Logistic Regression | Tuberculosis | 38 | 82% | 95% | 72% |
| Hee et al. 2019 [30] | GMM-UBM | Asthma | 176 | 91% | 82.81% | 84.76% |
| Porter et al. 2019 [31] | Neural Network | Croup, pneumonia, asthma, LRTD, bronchiolitis | 585 | Asthma:97% <br> LRTD:83, 82% <br> Bronchiolitis:84, 81% | Asthma: 91% <br> LRTD:– <br> Bronchiolitis:– | Asthma: – <br> LRTD: – <br> Bronchiolitis –: |
| Agbley et al. 2020 [171] | CNN | Covid-19 | COUGHVID | Cough:80% <br> – <br> – <br> – | Cough: 82.6% <br> Covid-19: 43% <br> Symptomatic:69% <br> Healthy: 51% | Cough:78.38% <br> Covid-19: 81% <br> Symptomatic:63% <br> Healthy 80% |
| Laguarta et al. 2020 [40] | CNN | Covid-19, alzheimer | 5320 | 98.5% | 98.5% | 94.2% |
| Bansal et al. 2020 [125] | CNN | Dry cough, wet cough, croup, pertussis and bronchitis,astma,COPD, Covid-19 | 501 | 70.58% | 80.95% | 64% |
| Pahar et al. 2020 [43] | LSTM | COVID-19, asthma, bronchitis | 1192 | 92.91% | 91% | 96% |
| Dunne et al. 2020 [128] | CNN | Covid-19 | 749 | 97.5% | – | – |
| Hassan et al. 2020 [126] | LSTM | Covid-19 | 80 | Breathing: 98.2% <br> Cough: 97% <br> Voice:88.2% | Cough: 98% | Cough: 96% |
| Danda et al. 2020 [127] | RF | Covid-19 | 3642 | Cough:94-96% <br> Sneeze:92–93% <br> Running Nose: 95-96% | Cough:94% <br> Sneeze:92% <br> Running Nose: 95% | – <br> – <br> – |
| Brown et al. 2020 [39] | LR | Covid-19, asthma | 6613 | Covid+/Covid-: 69% <br> With cough: 72% <br> Covid+/ asthma cough: 69% | – <br> – <br> – | |
| Balamurali et al. 2020 [172] | GMM-UBM | Asthma | 2362 | Asthma: 95% <br> Healthy: 89.9% | Overall: 95.6% <br> – | Overall: 95% <br> – |
| Bagad et al. 2020 [218] | CNN | Covid-19 | 3621 | – | 90% | 31% |
| Pal et al. 2020 [219] | DNN | Covid-19, asthma, bronchitis | 150 | Covid+: 96.81% <br> Covid-: 96.81% <br> Bronchitis: 93.46% <br> Asthma: 93.34% | Covid-19+: 91.39% <br> Covid-19-: 89.41% <br> Bronchitis: 88.45% <br> Asthma: 94.41% | Covid-19+: 97.49% <br> Covid-19-: 98.64% <br> Bronchitis: 98.08% <br> Asthma: 93.10% |
| Mouawad et al. 2021 [194] | XGBoost | Covid-19 | 1927 | 97% <br> 99% | 65% <br> 70% | – <br> – |
| Swarnkar et al. 2021 [220] | Logistic Regression | Croup, pneumonia | 224 | Croup:85, 82% <br> Pneumonia: <br> Breathing Index:76.67% <br> BI + cough: 95% | Croup: <br> Pneumonia:87, 85% <br> Breathing Index: 69.56% <br> BI + cough: 91.3% | Croup: <br> Pneumonia: <br> Breathing Index: 81.08% <br> BI + cough: 97.3 |
| Nessiem et al. 2021 [221] | CNN | COVID-19, asthma | 1427 | 67.7% | 77.6% | – |

TABLE VIII CONT. SUMMARY OF THE LITERATURE ON COUGH DIAGNOSIS ALGORITHMS.

| Author | Best AI Tech | Disease Diagnose | Sample size | Accuracy | Sensitivity | Specificity |
|---|---|---|---|---|---|---|
| Manshouri et al. 2021 [222] | SVM | Covid-19 | 16 | 94.21% | 89.58% | 97.26% |
| Pahar et al. 2021 [161] | Logistic Regression | Tuberculosis | 49 | – | TB: 93% | TB: 95% |
| Chowdhury et al. 2021 [42] | CNN | Covid-19 | 723 | Covid-19: 95.86%<br>Healthy:95.86%<br>Symptomatic (Covid-19):81.76%<br>Symptomatic (Healthy): 81.76% | Covid-19: 91.49%<br>Healthy: 97.8%<br>Symptomatic (Covid-19): 77.78%<br>Symptomatic (Healthy): 82.58% | Covid-19: 97.8%<br>Healthy:91.49%<br>Symptomatic (Covid-19):82.58%<br>Symptomatic (Healthy): 77.78% |
| Kumar et al. 2021 [223] | DCNN | Covid-19, asthma, bronchitis, pertussis | 1187 | Covid+: 93.57%<br>Pertussis: 93.86%<br>Bronchitis: 94%<br>Healthy: 95.45%<br>Asthma: 93.43% | Covid+: 95.45%<br>Pertussis: 93.57%<br>Bronchitis: 93.86%<br>Healthy: 94%<br>Asthma: 93.43% | Covid+: –<br>Covid-: –<br>–<br>–<br>– |
| Rao et al. 2021 [45] | DL | Covid-19 | DiCOVA 2021 dataset, COUGHVID | – | 80.49% | 77.88% |
| Mohammed et al. 2021 [224] | Spiking NN | Covid-19 | Coswara | 71% | 68% | 74% |
| Erdoğan et al 2021 [44] | SVM | Covid-19 | 1187 | 98.4 % | – | 97.3% |

TABLE IX. SUMMARY OF LITERATURE ON COUGH DETECTION AND DIAGNOSIS ALGORITHMS

| Author | Best AI Tech | Disease Diagnose | Sample size | Diagnosis Accuracy | Diagnosis Sensitivity | Diagnosis Specificity | Detection Accuracy | Detection Sensitivity | Detection Specificity |
|---|---|---|---|---|---|---|---|---|---|
| Pramono et al. 2016 [27] | LRM | Pertussis, croup, asthma, bronchiolitis | 38 | – | 92.38% | 90% | 85% | 85.09% | 98.32% |
| Imran et al. 2020 [32] | CNN+SVM | Covid-19, bronchitis, pertussis | 543 | 92.64 | Covid-19: 89.14%<br>Healthy: 94%<br>Bronchitis:93.86%<br>Pertussis: 93.57% | Covid-19: 96.67%<br>Healthy 99%<br>Bronchitis:96.33%<br>Pertussis:98.19% | 95.6% | 96.01% | 95.16% |
| Wei et al. 2020 [148] | CNN+SVM | Covid-19, bronchitis, chronis phyringitis, pertussis | 194 | covid-19: 76% | Bronchitis: 94.3%<br>Pharyngitis:85.3%<br>COVID-19:98.7%<br>Pertussis:99.8%<br>Healthy:96.3% | Bronchitis: 91.2%<br>Pharyngitis:83.3%<br>COVID-19:94.7%<br>Pertussis:95.8%<br>Healthy:92.1% | 76% | 98.8% | – |
| Bales et al. 2020 [24] | CNN | Bronchitis, bronchiolitis and pertussis | 268 | 89.6% | Pertussis:95%<br>Bronchitis:93.8%<br>Bronchiolitis:80% | Pertussis:96.9%<br>Bronchitis:87.5%<br>Bronchiolitis:100% | 89.05% | 91.901% | 86.2% |

Neighbor (k-NN), Decision Tree (DT), and Discriminant Analysis (DA) models. In online unit, the trained model is used to categorize the normal and COPD breath sounds. Classification based on the spirometry parameters and respiratory sound parameters is performed.

- *Snoring sounds*: Multiple studies performed research on the detection, classification, and analysis of the snoring sounds that can eventually be used for the diagnosis of obstructive sleep apnea. Sola et al. [19] trained an LR model using time and frequency parameters to track the respiratory disturbance index. The authors in [231] transformed the audio snoring signals to images and performed CNN classification. Moreover, to classify snore and non-snore events, the authors in [232] proposed an automatic and unsupervised detection framework. They recorded the respiratory sound signals from the patients and deployed fuzzy C-means clustering algorithm to achieve an accuracy of 98.6%. An AdaBoost classifier is used for the discrimination of snore and non-snore acoustic events in [233].

- *Auscultations sounds*: RespiratoryDatabase@TR database which contains lung auscultations at specific points is exploited in [234]. Several features are used to train Deep Belief Network (DBN) along with other classifiers such as SVM, k-NN, and Decision trees for comparison. DBN demonstrated the best performance in terms of sensitivity of 91%, accuracy of 93.67%, and specificity of 96.33%. Auscultations sounds are used for the diagnosis of pneumonia in [235]. The pre-processing of the data is performed using a unique method empirical mode decomposition. The data is used to extract the features in order to train SVM and k-NN for the prediction of pneumonia or non-pneumonia with an accuracy of 99.7%. In [22], a deep learning framework is employed that originally integrates MFCC-based pre-processing of auscultation audio data, advanced LSTM and Gated Recurrent Unit (GRU) models for the detection of respiratory abnormal sounds and of chronic/non-chronic diseases. Finally, from the extracted features, pathology driven prediction task is addressed along with the anomaly prediction task.

## VII. CHALLENGES AND FUTURE DIRECTIONS

AI algorithms are entitled to have acclaimed performance in terms of accuracy for image-based and sound-based detection/diagnosis methods. It is evident that AI has an immense potential to revolutionize the healthcare sector. The research community is continuously paving the way for the futuristic cost-effective, easy to use, and robust healthcare solutions. However, there are still some open research challenges that need to be addressed for further enhancement and innovations. This section briefly mentions some of the open challenges as well as the future directions to revitalize AI-based healthcare sector. These can be broadly classified into five major categories and are related to: 1) data, 2) model, 3) hardware/software resources, 4) privacy, and 5) interdisciplinary conglomeration. A summary of these categories is also provided in Fig. 12.

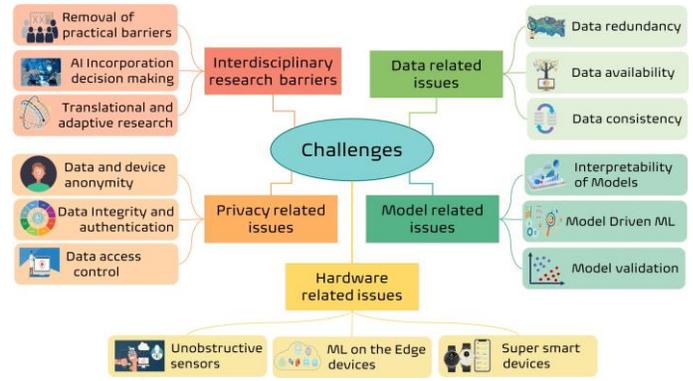

Fig. 12: Challenges faced during the incorporation of the AI techniques in healthcare domain.

### A. Data Collection, Consistency, and Availability

For the successful operation of any AI algorithm, extensive correctly labeled and representative data is required. From the existing literature it is evident that the available data is sparse and dedicated efforts are required for the collection of abundant representative data that is not redundant and is available to everyone. To address the first and foremost challenge, i.e., data collection and its availability, some researchers have collected data in the past and made it publicly available. Unfortunately there is a lot of redundancy in data and since there are no universal guidelines for the data acquisition, the collected data is inconsistent. In the past, several solutions were provided to deal with the missing data including self-organization maps and multiple imputation [236], however, active solutions are still required. In order to solve these issues, guidelines for data collection need to be defined. Moreover, to reduce biasness, data need to be collected with different and varying environmental setting and demographics. Since it is also important to correctly label the collected data, trained medical professionals are required to take part in this campaign. Finally, after successful collection, it is essential to make it open source so that the researchers across the globe could cross-validate their training models and thus develop robust and accurate detection/diagnosis algorithms. This would enable the research community and the healthcare sector to expedite the process of dealing with pandemics like COVID-19 in terms of capturing the underlying dynamics and uncertainties in a timely manner.

### B. AI-based Modeling and Domain Knowledge

Once a sufficient amount of correctly labeled data is available, the next challenge lies to train the optimal AI models. Since the targeted domain here is healthcare, the AI models should be highly accurate along with the better understanding of the underlying dynamics. The advancements in the learning process have undoubtedly improved the performance, however, it also increases the model complexity exceptionally by deploying it as a black-box. . This hinders to understand the insights about the learning behaviour, hence, leading to uninterpretable predictions.

For the AI models to be incorporated in the medical practice, it is highly anticipated that the models should be interpretable, accountable, and justify the prediction fairness across model [237]. Moreover, the outcome of the algorithms should be corroborated with medical professionals by providing the intuitive explanation and incorporating their domain knowledge. Thus, the developed model should be able to automate certain medical processes and significantly improve patient's health in the absence of medical personnel. Also, we should beware from the caveat that big data is always informative and the model is useful only when it is trained by the representative data. Hence, it is necessary to combine the data-driven methods with model-based methods.

### C. Hardware and Software Resources

After training the optimal model, automated screening tools are required that are capable of recording 24/7 and able to process the recorded information quickly to provide results promptly on demand. In addition, the tools should have the characteristics such as unobtrusiveness, compactness, privacy preservation, ability to suppress the ambient noise, and separate the cough from similar other sounds like speech, laughing, wheezing, and sneezing. Finally, they should be able to capture the variability of the cough acoustics within and between individuals, combined with the additional complexity of different respiratory diseases.

Utilization of smart portable devices and conglomeration of cloud computing is a viable solution to address these needs. Therefore, it is required to enable ML modeling on the edge devices and building a distributed intelligence system. This research in the domain of collaborative computational edge/fog computing is still in its infancy and more efforts are required to use the resources to improve the QoS of the end-to-end system.

### D. Privacy and Security

When the users provide data for the disease detection/diagnosis, it is important to forward and process the data securely with integrity and confidentially protected from eavesdropping. As adversary may use the data for illegal purposes, highlighted by many studies [123], [130], [170]. Transferring data to the cloud while using semi-automated systems can cause data breaching. Well known features, such as MFCC, used for the cough-based ML model training, lack the potential of keeping the patient data anonymous. Therefore, preserving the anonymity of data in the presence of power loss, failures, or attacks along with model security is a crucial aspect that needs to be incorporated in futuristic healthcare solutions [238].

To hide the patient identity, one solution is to only send weights to the cloud server rather than features such as MFCC [239]. Along with keeping the patient anonymity, data integrity services needs to be implemented to guarantee at the recipient end that the data has not been altered in transit by an adversary. Furthermore, adding authentication for both medical and non-medical applications can help verifying the data origin. Finally, since healthcare professionals are often also involved with the patient's physiological data, it is highly desirable that a role-based access control mechanism should be implemented in real-time healthcare applications that can restrict the access of the private physiological information.

### E. Interdisciplinary Research

In recent years, techniques have been evolved to become more refined and automated for the efficient information extraction from imaging and voice modalities by incorporating AI in order to ensure better patient care. However, most of the AI-based research is done by the researchers from technical sciences, and to deal with healthcare applications, specialization in bioinformatics, medical imaging, virology, and other related fields is also required. This challenge can be addressed either by collaborative work between experts from various related fields or by introducing new interdisciplinary specialization courses.

### F. Further Recommendations and Discussion

The application of AI for detecting and diagnosing the respiratory diseases has created an auspicious trend and myriad of future possibilities in the domain of healthcare. To overcome the aforementioned issues, there is a growing interest in the research community for devising frameworks and solutions by exploiting the advances of ubiquitous computing. By formulating the diagnostic inferencing issue in a form of sequential decision making process, that is backed up by the additional evidence generated from the relevant external resources, it is possible to develop a robust end-to-end disease diagnosis system. In addition, transfer learning can be exploited to accommodate the shortcomings of the limited data or unavailability of the computational resources.

However, we anticipate more research in the domain of deep learning for better segmentation and accurate detection of cough signatures. For DNN, creation of sufficient well-annotated and representative data sets are needed to accelerate the diagnosis process. There is also a need to make the dataset publicly available for research purpose and to develop the standardized formats. With the current advancement in technology and extensive recent literature detecting the onset of a various respiratory disease using smartphones, it can be inferred that such devices powered by AI, can be a suitable assistant to the medical practitioners. Such AI-based holistic systems have the potential to make healthcare more approachable and cost-effective. We can also leverage this fact for coping with the current pandemic like COVID-19 [32], [40], [43], [126], [145], [240]. The final frontier for making this solution more pragmatic is leveraging big data that would integrate technology, data mining for discovering patterns, clinical inference, and decision making under one platform [241]. Deep learning can explore and perceive the significant underlying quantitative biomarkers in Big data, that can assist the physicians for developing personalized treatment and management strategies. Lastly, necessary measures are required to motivate the healthcare community to incorporate the AI-based assistance for the accurate diagnosis. Hence, it is to be kept in mind, that the data-driven AI-based algorithms have the promising potential as an assistive tool for the preliminary detection and diagnosis in the realm of probabilities, however, the final verdict will still lie with the medical practitioners.

## VIII. Conclusions

Recent advances in the healthcare domain demand revolutionized practices. There is an unprecedented interest towards data-driven processes to unleash the computing power that AI can provide. Machine Learning-based frameworks are being leveraged for the general diagnosis of virulent maladies. By gathering these studies, this survey provides a comprehensive study on the existing literature on detection and preliminary diagnosis of the respiratory diseases with the aid of cough sounds and AI-based models. Moreover, this survey presented methodologies, data collection procedures, and analyzed objective assessment algorithms, that are employed in the reviewed studies. Additionally, it also analyzed the studies in the broad categories, i.e., detection and diagnosis using cough acoustic and then by using similar sounds (lungs sounds, breathing, auscultations, and snoring). Lastly, it discussed several challenges and vulnerabilities that need to be addressed for the successful formulation of the ML pipeline in the healthcare along with the potential future extensions. The study concludes that AI-powered solutions demonstrate promising potential for developing innovative clinical decision assistance and diagnostic tools that can help the healthcare community and policy makers to revitalize the existing healthcare practices.